\documentclass[amsmath,amssymb, aps,prx,reprint,nofootinbib,superscriptaddress]{revtex4-1}

\usepackage{float}
\usepackage{amsmath,amssymb}
\usepackage[english]{babel}
\usepackage{upgreek}
\usepackage{scrextend}
\usepackage{dsfont}
\usepackage{graphicx}
\usepackage{graphicx}
\usepackage{color}
\usepackage{bbold}

\DeclareMathOperator\erf{erf}

\begin{document}
\title{Macroscopicity of quantum mechanical superposition tests via hypothesis falsification}
\author{Bj\"{o}rn Schrinski}
\affiliation{
 University of Duisburg-Essen, Faculty of Physics, Lotharstra\ss e 1, 47048 Duisburg, Germany}
\author{Stefan Nimmrichter}
\affiliation{
Max Planck Institute for the Science of Light, Staudtstra{\ss}e 2, 91058 Erlangen}
\author{Benjamin A. Stickler}
\affiliation{
 University of Duisburg-Essen, Faculty of Physics, Lotharstra\ss e 1, 47048 Duisburg, Germany}
\affiliation{
 QOLS, Blackett Laboratory, Imperial College London, London SW7 2BW,
United Kingdom }
\author{Klaus Hornberger}
\affiliation{
 University of Duisburg-Essen, Faculty of Physics, Lotharstra\ss e 1, 47048 Duisburg, Germany}

\begin{abstract}
We establish an objective scheme to determine the macroscopicity of 
quantum superposition tests with mechanical degrees of freedom. It is based on the Bayesian hypothesis falsification of a class of macrorealist modifications of quantum theory, such as the model of Continuous Spontaneous Localization. The measure uses the raw data gathered in an experiment, taking into account all measurement uncertainties, and can be used to directly assess any conceivable quantum mechanical test.
We determine the resulting macroscopicity for three recent tests of quantum physics: double-well interference of Bose-Einstein condensates, Leggett-Garg tests with atomic random walks, and entanglement generation and read-out of nanomechanical oscillators.
\end{abstract}

\maketitle
\section{Introduction}

Any experiment witnessing or exploiting quantum coherent phenomena may be viewed as a test of whether quantum theory is complete at a fundamental level. While quantum mechanics is supported by all empirical observations up to date, all these observations are equally compatible with a number of alternative theories restoring macroscopic realism and resolving the measurement problem  \cite{Leggett2002,Bassi2013}.

In recent years, various experiments demonstrated quantum superpositions or entanglement with mechanical objects of increasingly high masses and particle number, involving ever larger spatial delocalizations and coherence times. They include setups as diverse as
counter-propagating superconducting loop currents \cite{friedman2000,vanDerWal2000},
large path-separation atom interferometers \cite{peters1999,dimopoulos2007},
high-mass molecular near-field interferometers \cite{Gerlich2011,eibenberger2013},
trapped and freely-falling Bose-Einstein condensates \cite{Schmiedmayer2013,Kovachy2015},
de-localized states and Leggett-Garg tests in optical lattices \cite{Alberti2009,Robens2015},
entangled ion chains \cite{jurcevic2014,islam2015}, and nanomechanical oscillators \cite{Riedinger2018,ockeloen2018,Marinkovi2018}. While all these experiments establish variants of a Schr{\"o}dinger-cat-like state, an  obvious question is the degree of macroscopicity (or `cattiness') reached.

There are many ways to assess the macroscopicity of a Schr{\"o}dinger cat realized in a quantum experiment \cite{frowis2018}. Most measures quantify the complexity of the quantum state based on information- or resource-theoretic concepts \cite{Korsbakken2007,Marquardt2008,Froewis2012,yadin2016general,Yadin2018}, or introduce suitable distance measures in Hilbert space \cite{Bjoerk2004,Lee2011}. While such abstract state vector ranking schemes may be used to compare experimental setups of similar kind, none can cover the entire variety of present-day superposition experiments \cite{friedman2000,vanDerWal2000,peters1999,dimopoulos2007,Gerlich2011,eibenberger2013,Schmiedmayer2013,Kovachy2015,Alberti2009,Robens2015,jurcevic2014,islam2015,Riedinger2018,ockeloen2018,Marinkovi2018}.

A viable alternative is to regard a Schr{\"o}dinger cat as more macroscopic than others if its demonstration is more at odds with the classical expectations shaped by our  every-day experiences. In Ref.~\cite{Nimmrichter2013} this was cast into a macroscopicity  measure by quantifying the extent to which a superposition experiment rules out a 
natural class of objective modifications of quantum theory that predict classical behavior on the macroscale.  A prominent example of such  \emph{classicalizing modifications} is the model of \emph{Continuous Spontaneous Localization} (CSL) \cite{Bassi2013}.
Recent tests of nonlocality and macrorealism, demonstrating the violation of Bell and Leggett-Garg  inequalities at unprecedented mass and time scales, call for a generalization of this measure for arbitrary quantum tests with mechanical degrees of freedom.

\begin{figure}
  \centering
  \includegraphics[width=0.39\textwidth]{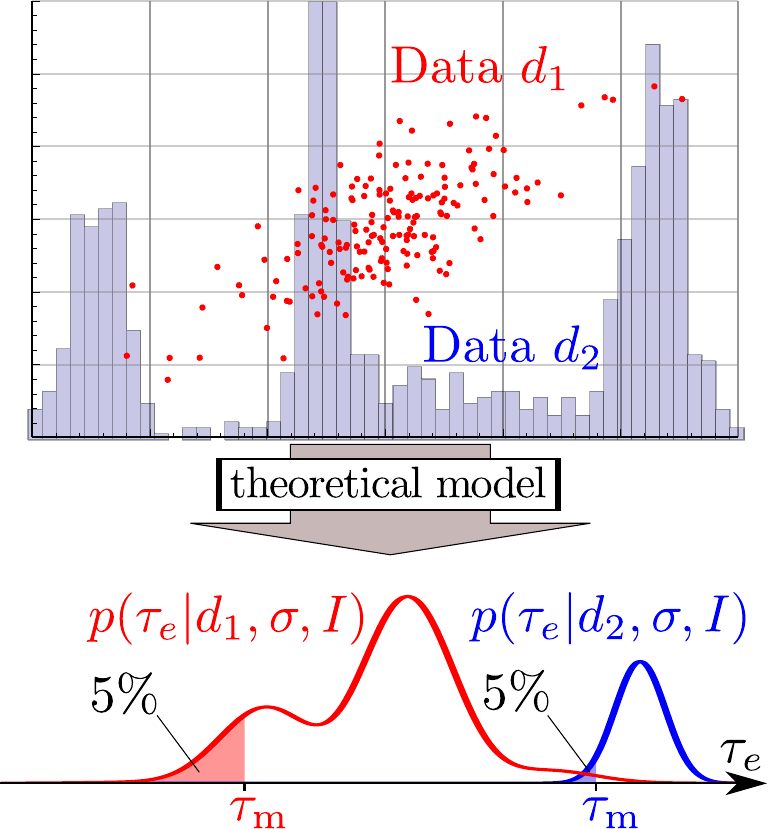}
  \caption{Scheme to compare the macroscopicity of two different quantum superposition tests: The experiments deliver raw data sets $d_1$ and $d_2$, which may be of arbitrary type and structure. They can be used to rule out  modifications  of standard quantum theory which classicalize the dynamics. Combining the data with the theoretical expectation yields a probability distribution for the classicalization timescale $\tau_e$, given the modification parameters $\sigma$ and the background information $I$. A quantum experiment is considered more macroscopic if the data rule out greater values of $\tau_e$, as inferred from the $5\,\%$ quantile $\tau_{\rm m}$.}
\label{fig:scheme}
\end{figure}

In this article, we present the most general framework for assigning the macroscopicity reached in quantum mechanical superposition experiments, based on non-informative Bayesian hypothesis testing, see Fig.~\ref{fig:scheme}.  As the natural generalization of the measure presented in \cite{Nimmrichter2013}, it relies only on the empirical evidence (i.e. the raw measurement outcomes)  delivered by a given superposition test. It thus accounts for the measurement imperfections independently of the chosen experimental figure of merit, such as the fringe visibility or an entanglement witness.

This measure of macroscopicity can be applied to assess any mechanical superposition experiment. It is unbiased by construction and it accounts naturally for experimental uncertainties and statistical fluctuations. These advantages come at the expense of a certain theoretical effort required for calculating the macroscopicity of a given experiment. Specifically, the time evolution of the quantum system must be calculated in presence of classicalizing modifications to obtain the probability distribution for all possible measurement outcomes. In the second part of this article we demonstrate how this task is accomplished for three superposition tests at the cutting-edge of quantum physics: double-well interference of number-squeezed Bose-Einstein condensates (BECs) \cite{Schmiedmayer2013}, Leggett-Garg inequality tests with atomic quantum random walks \cite{Robens2015}, and generation and witnessing of entanglement between two spatially separated nanomechanical oscillators \cite{Riedinger2018}.

\begin{figure*}
  \centering
  \includegraphics[width=0.99\textwidth]{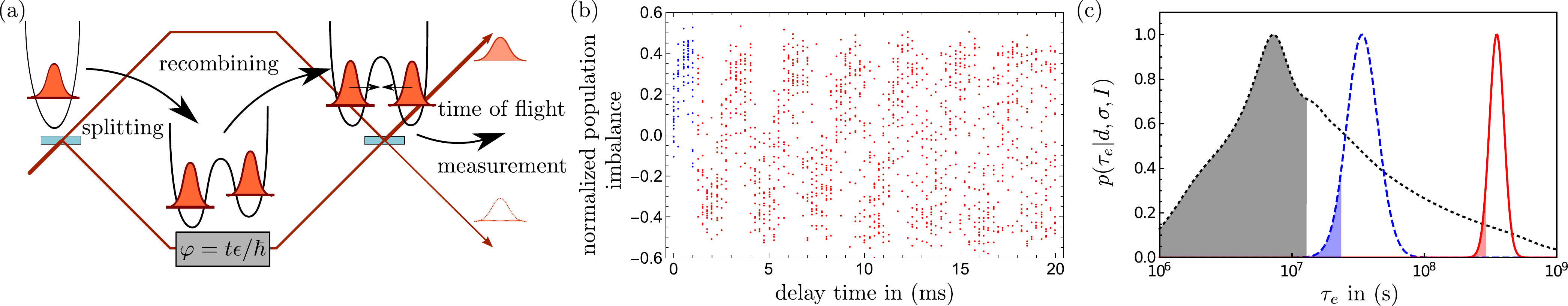}
  \caption{(a) Schematic illustration of the double-well BEC interference experiment \cite{Schmiedmayer2013}. The BEC is initially split into a superposition between slightly detuned left and right double-well states, then number squeezed, then let to freely evolve for a delay time, before a final $\pi/2$-pulse (recombiner) converts the phase difference between the states into an occupation difference. (b) Time-of-flight measurement data of the occupation imbalance versus delay time (from Ref.~\cite{Schmiedmayer2013}). (c) Posterior distribution of the classicalization timescale (red solid line)  as obtained via Bayesian updating of Jeffreys' prior (black dashed line) with the measurement data. The blue line is the intermediate distribution obtained by using only the blue data points up to one millisecond in (b). The shaded areas indicate the lowest five percent quantiles and all distributions are normalized to the same maximum value.}
\label{fig:2}
\end{figure*}
\begin{figure*}
  \centering
  \includegraphics[width=0.99\textwidth]{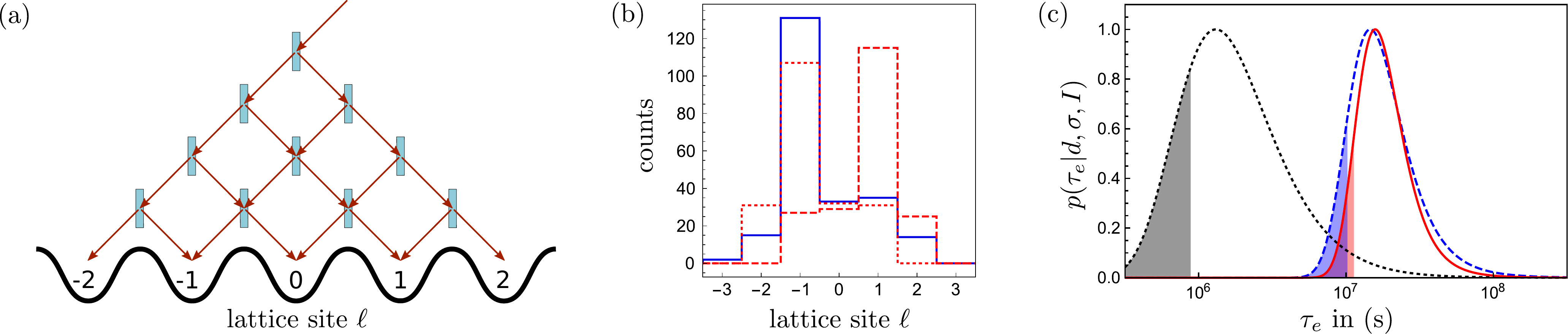}
  \caption{(a) Schematic illustration of the quantum random walk consisting of four consecutive steps. In each step the atom is coherently split into a left- and right-moving state, and the final populations are read-out after the fourth step. (b) Experimental data from Ref.\,\cite{Robens2015}. The blue solid line is the data from the total quantum random walk, while the red lines are conditioned on the first step being either left (dashed) or right (dotted). (c) Posterior distribution of the classicalization timescale (red solid line) as obtained via Bayesian updating of Jeffreys' prior (black dashed line) with the measurement data. The blue line is the intermediate distribution obtained by using only the blue measurement runs in (b). The shaded areas indicate the lowest five percent quantiles and all distributions are normalized to the same maximum value.}
\label{fig:3}
\end{figure*}
\begin{figure*}
  \centering
  \includegraphics[width=0.99\textwidth]{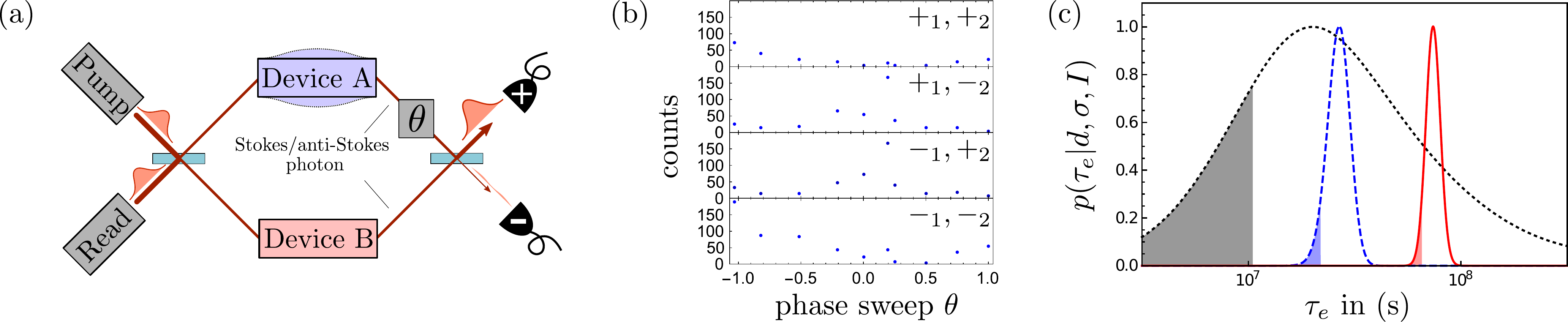}
  \caption{(a) Stoke's scattering of a pump photon, prepared in a spatial superposition by the entrance beam splitter, generates entanglement between two nanomechanical oscillators, which is then read-out by anti-Stoke's scattering of a read photon. Entanglement is certified by a coincidence measurement of the scattered photons in the upper ($+$) or lower ($-$) detector behind the exit beam splitter. (b) Measurement data \cite{Riedinger2018} as a function of the tunable relative phase $\theta$  between the two interferometer arms ({\it phase sweep}). (c) Posterior distribution of the classicalization timescale (red solid line) as obtained via Bayesian updating of Jeffreys' prior (black dashed line) with the measurement data. The blue line is the posterior obtained by taking only phase sweep data points into account, while the red line also accounts for measurements with variable time delay between pump and read [{\it time sweep}; not shown in (b)]. The shaded areas indicate the lowest five percent quantiles and all distributions are normalized to the same maximum value.}
\label{fig:4}
\end{figure*}

\section{Macroscopicity of three recent superposition tests} \label{sec:neue2}

Before presenting the formal framework of the proposed measure of macroscopicity, we illustrate its application to three recent superposition tests \cite{Schmiedmayer2013,Robens2015,Riedinger2018}.  As a common theme, these experiments use derived quantities, such as visibilities, correlation functions, and entanglement witnesses, to certify the quantumness of their observations. One important advantage of the Bayesian approach advocated here is that it is independent of such data processing (and thus of secondary observables) and based exclusively on likelihoods associated with elementary measurement events. A theoretical derivation of the likelihoods required to assess the three mentioned experiments is presented in Secs. \ref{sec:3}--\ref{sec:5}.

The measure uses the experimental data $d$ to determine the posterior probability distribution $p(\tau_e|d,\sigma, I)$ of classicalization timescales $\tau_e$, given  the modification parameters $\sigma$, and any background information $I$ required to model the experiment. To ensure that each experiment is rated without bias, the least informative prior is used for Bayesian updating to yield the final posterior distribution. Figures \ref{fig:2}--\ref{fig:4} show how disparate experimental measurement protocols and data sets \cite{Schmiedmayer2013,Robens2015,Riedinger2018} yield comparable posterior distributions, narrowly peaked around a definite modification timescale. 
As an increasing number of data-points is included in the Bayesian updating procedure, the distributions shift to higher modification time scales, while their widths decrease. The lowest five percent quantile $\tau_{\rm m}(\sigma)$ of the posterior distribution determines the macroscopicity as
\begin{equation*}
\mu_{\rm m} = \max_\sigma \left [ \log_{10}\left(\frac{\tau_{\rm m}(\sigma)}{1\,{\rm s}}\right) \right ].
\end{equation*}
The value $\mu_{\rm m}$ thus quantifies the degree to which the quantum measurement data rules out a 
natural class of classicalizing modifications of quantum theory. 

The resulting macroscopicities of the experiments are: $\mu_{\rm m}=8.5$ for the BEC interferometer \cite{Schmiedmayer2013}, $\mu_{\rm m}=7.1$ for the atomic Leggett-Garg test \cite{Robens2015}, and $\mu_{\rm m}=7.8$ for the entangled nanobeams. That the BEC and the atomic random walk experiments exhibit comparable macroscopicities is due to the fact that they both witness single atom interference at a similar product of squared mass and coherence time.  The macroscopicity associated with the entangled nanobeam experiment is roughly on the same order of magnitude on the logarithmic scale, despite the high mass and the large separation between the two beams and as well as coherence times of microseconds. This surprising result can be explained by the fact that the probed superposition state is delocalized merely by a few femtometers, and thus probes quantum theory only on sub-atomic scales.

Comparison of the three experiments also reveals that the convergence rate of the posterior distribution can  vary strongly. In case of the Leggett-Garg test with an atomic quantum random walk \cite{Robens2015}, the data set consists of 627 walks which all end in one of five final lattice sites. Since the likelihood of two of those outcomes is independent of the modification they include no information for the hypothesis test, which slows the convergence of the Bayesian updating procedure. In contrast, the double-well BEC-interferometer \cite{Schmiedmayer2013}  provides a distribution of measurement outcomes over a practically continuous range of values, so that each experimental run yields a high degree of information gain, implying that 1457 measured population imbalances lead to a relatively narrow posterior distribution. In the case of nanobeams only two of four possible coincidence outcomes have different likelihoods, and thus several thousand repetitions of the measurement protocol are required to make the posterior converge.

\section{Macroscopicity via hypothesis falsification} \label{sec:2}

\subsection{Empirical measure of macroscopicity}\label{sec:2A}

Classicalizing modifications of quantum theory propose an alternative (stochastic) evolution equation for the wavefunction. The observable consequences of these alternative theories are then encoded in the dynamics of the state operator $\rho_t$, which evolves according to a modified von Neumann equation
\begin{align} \label{eq:modvonneum}
\partial_t \rho_t=\mathcal{L}\rho_t+\frac{1}{\tau_e}\mathcal{M}_{\sigma}\rho_t.
\end{align}
Here ${\cal L}\rho_t$ denotes the time evolution according to standard quantum theory (including possible decoherence) and ${\cal M}_\sigma \rho_t / \tau_e$ describes the effect of the proposed modification, characterized by the time scale $\tau_e$ and the set of modification parameters $\sigma$.

Indeed, a generic class of modification theories are compatible with all observations up to date, and they restore realism on the macroscale. This class can be parametrized by imposing a few natural consistency requirements, such as Galilean 
invariance and exchange symmetry \cite{Nimmrichter2013}. The parameters $\sigma=(\sigma_q,\sigma_s)$ with the  dimensions of momentum and length, respectively, then specify the length and momentum scale on which the modification acts by means of the distribution function $g_\sigma(q,s)$ with zero mean and widths $\sigma_q,\sigma_s$,
\begin{align}
\mathcal{M}_{\sigma}\rho_t=&\int d^3{\bf q}d^3{\bf s}\,g_\sigma(q,s)\left [
{\sf L}({\bf q},{\bf s})\rho_t{\sf L}^{\dagger}({\bf q},{\bf s}) \vphantom{\frac{1}{2}}\right. \nonumber \\
& \left. -\frac{1}{2}\left\{{\sf L}^{\dagger}({\bf q},{\bf s}){\sf L}({\bf q},{\bf s}),\rho_t
\right\}\right].
\label{eq:MIMpointparticle}
\end{align}
The Lindblad operators in second quantization,
\begin{align} \label{eq:lindblad}
{\sf L}({\bf q},{\bf s})=\sum_\alpha\frac{m_\alpha}{m_e}\int d^3{\bf p}\,e^{i{\bf p}\cdot m_e{\bf s}/m_\alpha\hbar}{\sf c}_\alpha^{\dagger}({\bf p}){\sf c}_\alpha({\bf p}-{\bf q})\,,
\end{align}
induce displacements in phase-space by means of the annihilation operator ${\sf c}_\alpha({\bf p})$ for momentum ${\bf p}$. They involve a sum over the different particle species $\alpha$ with mass $m_\alpha$, whose ratio over the electron mass $m_e$  effectively amplifies the strength of the modification for heavy particles, ensuring that macrorealism is restored \cite{Nimmrichter2013}. 

Roughly speaking, phase-space superpositions of a particle of mass $m_\alpha$ will decohere at the maximal amplified rate ($m_\alpha/m_e)^2/\tau_e$ if they extend over spatial distances greater than $\hbar/\sigma_q$ or momentum distances greater than $m_\alpha \hbar / m_e \sigma_s$. 
We take $g_\sigma$ to be Gaussian in the following. The modification \eqref{eq:MIMpointparticle} then reduces to the model of CSL \cite{Bassi2013} for fixed $\sigma_q$ and $\sigma_s=0$. As explained in Ref.~\cite{Nimmrichter2013}, the bounds $\hbar/\sigma_q\gtrsim  10\,$fm and $\sigma_s\lesssim 20\,$pm ensure that the modification does not drive the system into the regime of relativistic quantum mechanics. In what follows, we will define the empirical measure of macroscopicity as the extent to which a quantum experiment rules out such classicalizing modifications.

Since the modified evolution \eqref{eq:modvonneum} predicts deviations from standard quantum mechanics at some scale these modification theories are empirically falsifiable. Thus, any quantum experiment gathering measurement data $d$ can be considered as testing the hypothesis $H_{\tau_e^*}$:
\begin{quote}
	\textit{Given a classicalizing modification \eqref{eq:MIMpointparticle} with parameters $\sigma$, the dynamics of the system state $\rho_t$ are determined by Eq.~\eqref{eq:modvonneum} with a modification time scale  $ \tau_e \leq \tau_e^*$.}
\end{quote}
Note that greater values of $\tau_e$ imply weaker modifications.

The empirical data $d$  determine the Bayesian probability $P(H_{\tau_e^*} | d,\sigma, I)$ that $H_{\tau_e^*}$ is true, given the background information $I$. The latter includes all knowledge required for describing the experiment, such as the Hamiltonian, environmental decoherence processes, and the measurement protocol.

In order to compare $H_{\tau_e^*}$ with the complementary hypothesis $\overline{H}_{\tau_e^*}$ that the modification time
scale $\tau_e$ is larger than $\tau_e^*$ (including unmodified quantum mechanics as $\tau_e=\infty$), one defines the odds ratio \cite{von2014bayesian}
\begin{equation} \label{eq:oddsratio1}
o(\tau_e^* | d,\sigma, I) = \frac{P(H_{\tau_e^*} | d,\sigma, I)}{P(\overline{H}_{\tau_e^*} | d,\sigma, I)}.
\end{equation}
If the data implies that the odds ratio is less than a certain maximally acceptable value $o_{\rm m}$ we can favor $\overline{H}_{\tau_e^*}$ over $H_{\tau_e^*}$. Modifications of quantum theory with $\tau_e \leq \tau_e^*$ are then ruled out by the data at odds $o_{\rm m}$ .

In order to evaluate the odds ratio \eqref{eq:oddsratio1} we use Bayes' theorem and exploit that for the hypothesis test to be unbiased by earlier experiments, $H_{\tau_e^*}$ and $\overline{H}_{\tau_e^*}$ must be {\em a priori} equally probable. Further using that the hypothesis $H_{\tau_e^*}$ implies $\tau_e \leq \tau_e^*$ yields 
\begin{align}
o(\tau_e^*|d,\sigma,I)=&\frac{\displaystyle \int_0^{\tau_e^*}d\tau_e\,P(d|\tau_e,\sigma,I)p(\tau_e|\sigma,I)}{\displaystyle \int_{\tau_e^*}^{\infty}d\tau_e\,P(d|\tau_e,\sigma,I)p(\tau_e|\sigma,I)},
\label{eq:oddsratio}
\end{align}
where $p(\tau_e|\sigma,I)$ is the prior distribution of $\tau_e$, whose choice will be discussed in Sec.~\ref{subsec:32}. The probabilities $P(d |\tau_e,\sigma,I)$ are independent of the hypothesis $H_{\tau_e^*}$; they can be calculated for any experiment by solving the modified evolution equation \eqref{eq:modvonneum} with classicalization time scale $\tau_e$ and parameters $\sigma$.

The data $d$ is usually gathered in $N$ consecutive independent runs, $d = \{d_1,d_2,\ldots,d_{N}\}$, where $d_k$ denotes the set of (possibly correlated) measurement outcomes of round $k$. The likelihood for the entire data set $d$ is then given by
\begin{align}
P(d|\tau_e,\sigma,I)=\prod_{k}P(d_k|\tau_e,\sigma,I).
\label{eq:likelihoodproduct}
\end{align}
Every additional experimental run thus refines the posterior probability density, according to Bayes' theorem
\begin{align} \label{eq:posterior}
p(\tau_e|d,\sigma,I) = \frac{P(d|\tau_e,\sigma,I)p(\tau_e|\sigma,I)}{P(d|\sigma,I)},
\end{align}
where the normalization constant $P(d|\sigma,I)$ plays no role for the odds ratio.
For sufficiently large data sets and for well behaved priors the posterior is independent of the prior distribution $p(\tau_e|\sigma, I)$  \cite{schwartz1965bayes,ghosh2003springer}.

For what follows, we choose the threshold odds $o_{\rm m} = 1:19$, corresponding to the posterior probability 
\begin{equation} \label{eq:taum}
P(\tau_e \leq \tau_{\rm m}|d,\sigma,I) \equiv \int_0^{\tau_{\rm m}} d \tau_e p(\tau_e \vert d,\sigma,I) = 5\,\%.
\end{equation}
This determines the greatest excluded modification time scale $\tau_{\rm m}$ (at odds $o_{\rm m}$) so that for all $\tau_e^* \leq \tau_{\rm m}$ the odds ratio \eqref{eq:oddsratio} is smaller than $o_{\rm m}$ for given modification parameters $\sigma$.

Given the greatest excluded modification time scale $\tau_{\rm m}(\sigma)$ as a function of the modification parameters $\sigma$, one defines the empirical measure of macroscopicity as
\begin{equation}
  \mu_{\rm m} = \max_\sigma \left [ \log_{10}\left(\frac{\tau_{\rm m}(\sigma)}{1\,{\rm s}}\right) \right ],
\label{eq:Macroscopicity}
 \end{equation}
where $\tau_{\rm m}(\sigma)$ [Eq.~\eqref{eq:taum}] is the extent to which the measurement data $d$ of a given quantum experiment rules out the class of modifications \eqref{eq:MIMpointparticle}. The value of $\mu_{\rm m}$ thus ranks superposition experiments against each other according to the degree to which they are at odds with our classical expectation. 

We emphasize that this definition must only be used for experiments that undeniably show genuine quantum signatures. It cannot be used to \emph{certify} whether a given experiment observes a superposition state. This is due to the fact that the absence of modification-induced heating and momentum diffusion can be observed also in classical experiments. Even though quantum coherence plays no role in such setups, they can serve to exclude combinations of classicalization timescales and modification parameters \cite{Laloe2014,Nimmrichter2014,carlesso2016experimental,li2016discriminating,
goldwater2016testing, vinante2017, schrinski2017collapse,adler2018bulk,bahrami2018testing}.

Even in genuine quantum superposition experiments the observed absence of modification-induced heating may dominate the range of excluded modification parameters. In this case it is necessary to recombine the observables in such a way that they separate into a subset of random variables $D$ providing information about quantum coherence and a subset  $D_{\rm heat}$ yielding only information about the energy gain. (For example, in the case of the double-well BEC interference experiment, where one measures the particle numbers in the two different wells, their difference shows interference based on quantum coherence, while their sum constraints particle loss due to heating.) For a fair assessment of the macroscopicity, the likelihood $P(D,d_{\rm heat}|\tau_e,\sigma, I)$ must be conditioned on the realized data $d_{\rm heat}$ restricting modification induced heating,
\begin{equation}\label{eq:10cl}
P(D|\tau_e,\sigma, I, d_{\rm heat}) = \frac{P(D,d_{\rm heat}|\tau_e,\sigma, I)}{ P(d_{\rm heat}|\tau_e,\sigma, I)}\,
\end{equation}
with $P(d_{\rm heat}|\tau_e,\sigma, I)=\sum_D P(D,d_{\rm heat}|\tau_e,\sigma, I)$. This way the witnessed lack of heating is effectively added to the background information $I$. (It also shows how to formally take into account the observation that the experiment could be executed at all, i.e.\ that the setup did not disintegrate due to modification-induced heating.) In Sect.~\ref{sec:3} we demonstrate how the conditioning on quantum observables works in practice by means of a nontrivial 	example.

\subsection{Jeffreys' prior} \label{subsec:32}

If the data set is not sufficiently large, the measure \eqref{eq:Macroscopicity} will in general depend on the prior distribution chosen to evaluate the odds ratio \eqref{eq:oddsratio}. It is therefore necessary to specify which prior distribution $p(\tau_e | \sigma,I)$ must be used to calculate the macroscopicity \eqref{eq:Macroscopicity}.

In order to ensure that the macroscopicity $\mu_{\rm m}$ does not have a bias towards a selected class of quantum superposition tests, the prior must be chosen in the most uninformative way, i.e.\ without including any {\it a priori} believes. For instance, this implies that it must not play a role whether we use the time scale $\tau_e$ or the rate $1/\tau_e$ to parametrize the class of modifications, which  already excludes a uniform or piecewise-constant prior. Therefore, the natural choice is Jeffreys' prior \cite{jeffreys1998theory}. Given the likelihood $P(d|\tau_e, \sigma,I)$ associated with a random variable $d$, it is defined as the square root of the Fisher information,
\begin{align}
p(\tau_e|\sigma,I)\propto&\sqrt{\mathcal{I}(\tau_e|\sigma,I)}\nonumber\\
=&\sqrt{ \left \langle \left  ( \frac{\partial}{\partial \tau_e}\log[P(D|\tau_e,\sigma,I)] \right )^2\right \rangle_D}\,.
\label{eq:JeffreysPrior}
\end{align} 
The ensemble average $\langle \cdot \rangle_D$ is performed over the entire range of possible measurement outcomes $D$ with Probability $P(D |\tau_e,\sigma,I)$.

This prior coincides with the so-called reference prior, so that it maximizes the Kullback-Leibler-divergence between prior and posterior and thus the average information gain in the Bayesian updating process \eqref{eq:posterior} \cite{bernardo1979reference,ghosh2011objective}. In this sense, Jeffreys' prior can be considered as the least informative prior \cite{berger2009formal}. In addition, it is invariant under re-parametrizations of the model \cite{jeffreys1998theory}, implying that it is irrelevant whether we use the timescale $\tau_e$ or the rate $\lambda=1/\tau_e$ (as employed in the model of Continuous Spontaneous Localization \cite{Bassi2013}) or any other power of $\tau_e$ as the fundamental parameter of our model.
We demonstrate in App.\,\ref{app:beweis} that for all practical purposes Eq.~\eqref{eq:JeffreysPrior} yields a normalizable posterior distribution \eqref{eq:posterior} because the master equation \eqref{eq:modvonneum} and thus the likelihood $P(d|\tau_e,\sigma, I)$ are smooth functions of $\tau_e$. 

If different measurement protocols are implemented, indicated here by the index $k$ (typical scenarios are different waiting times in a time integrated interferometer), Jeffreys' prior is weighted as
\begin{align}\label{eq:JeffreysPriorWeighted}
p(\tau_e|\sigma,I)\propto\sqrt{\sum_k N_k\,\mathcal{I}(\tau_e|\sigma,I_k)}.
\end{align}
Here, $N_k$ is the number of experimental runs with the respective $P(D_k|\tau_e,\sigma,I_k)$. The simple form of Jeffreys' prior \eqref{eq:JeffreysPriorWeighted} can be obtained by noting that 
$\langle \partial_{\tau_e}\log[P(D_k|\tau_e,\sigma,I_k)] \rangle_{D_k}=0$ in any case since the normalization of the probability distribution $P(D_k|\tau_e,\sigma,I_k)$ must be preserved for all $\tau_e$.

\subsection{General scheme for assigning macroscopicities}

The formal framework of how to assess the macroscopicity of arbitrary quantum mechanical superposition tests is now complete:
\begin{enumerate}
	\item Determine the Hamiltonian, environmental decoherence channels, and quantum measurement protocol, and use these to calculate the likelihood $P(D|\tau_e, \sigma,I)$ in presence of the modification \eqref{eq:MIMpointparticle}. If appropriate use Eq.~\eqref{eq:10cl} to focus on data demonstrating quantum coherence.
	\item Calculate Jeffreys' prior \eqref{eq:JeffreysPriorWeighted}.
	\item Determine the posterior distribution via Bayesian updating \eqref{eq:posterior} to extract $\tau_{\rm m}(\sigma)$ via \eqref{eq:taum}.
	\item Find the maximum of the function $\tau_{\rm m}(\sigma)$, which determines the macroscopicity \eqref{eq:Macroscopicity}.
\end{enumerate}

This recipe prescribes how to calculate the macroscopicity  based on the empirical evidence of a quantum experiment. It formalizes and generalizes the notion of macroscopicity introduced in Ref.~\cite{Nimmrichter2013}. The approximate expressions derived in Ref.~\cite{Nimmrichter2013} intrinsically assume that imperfections of a given experiment yield a  definite value of $\tau_e<\infty$, corresponding to a delta-peaked posterior distribution. The Bayesian framework put forward here extends this to measurement schemes and data sets yielding a finite posterior distribution $p(\tau_e|d,\sigma,I)$. It is thus the natural extension for noisy data and arbitrary measurement strategies.

In practice, the most complicated part of the above scheme is calculating the likelihoods in step 1. This requires finding an appropriate and quantitative description of the quantum dynamics  in presence of decoherence and the modification. Note that the macroscopicity is underestimated if relevant decoherence channels are neglected in the calculation of the likelihoods. The remainder of this article demonstrates how the likelihoods can be calculated for the three superposition tests discussed in Sec.~\ref{sec:neue2}.

\section{Ramsey interferometry with a number-squeezed BEC} \label{sec:3}

\subsection{Experimental Setting and Basics}

In the experiment reported in Ref.~\cite{Schmiedmayer2013} a ${}^{87}$Rb BEC is trapped in a double-well potential and made to interfere, see Fig.~\ref{fig:2}(a). The two involved modes  $a,b$ form an effective two-level system described by the annihilation  operators ${\sf c}_{a}$, ${\sf c}_{b}$. The state of the BEC can thus be represented by a collective pseudospin, defined by means of the (dimensionless) quasi angular momentum operators \cite{Arecchi1972}
\begin{align}
{\sf J}_x&=\frac{1}{2}\left({\sf c}^{\dagger}_a{\sf c}_b+{\sf c}^{\dagger}_b{\sf c}_a\right)\nonumber\\
{\sf J}_y&=\frac{1}{2i}\left({\sf c}^{\dagger}_a{\sf c}_b-{\sf c}^{\dagger}_b{\sf c}_a\right)\nonumber\\
{\sf J}_z&=\frac{1}{2}\left({\sf c}^{\dagger}_a{\sf c}_a - {\sf c}^{\dagger}_b{\sf c}_b\right).
\label{eq:AngularMomentumOperators}
\end{align}
They fulfill the angular momentum commutation relations $[{\sf J}_{\lambda},{\sf J}_{\mu}]=i\epsilon_{\lambda,\mu,\nu}{\sf J}_{\nu}$. The simultaneous eigenstates of ${\sf J}^2$ with eigenvalue $J(J+1)$ and ${\sf J}_z$ with eigenvalue $m$ are denoted by $\left|J,m\right\rangle$ (Dicke state), where $J=N/2$.

\begin{figure*}
  \centering
\includegraphics[width=0.99\textwidth]{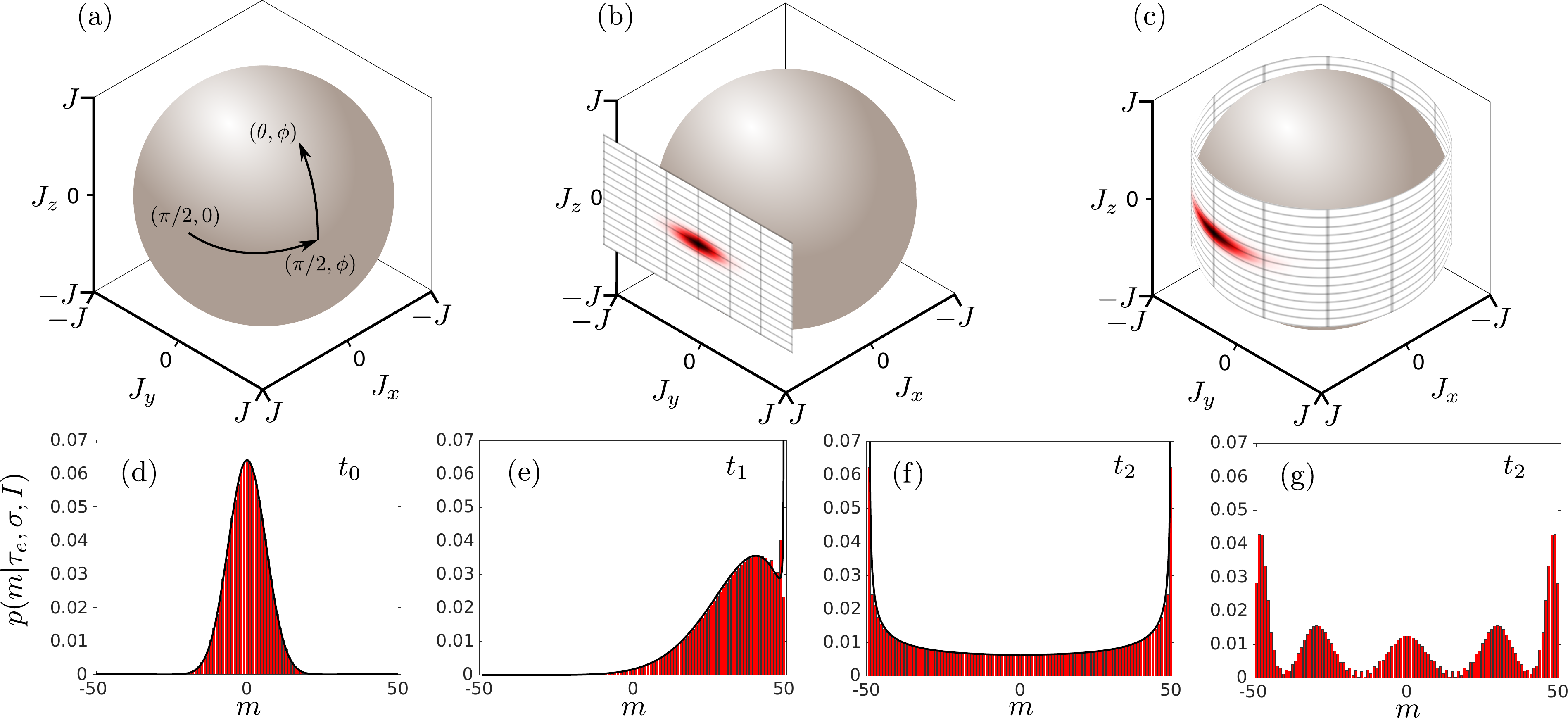}
  \caption{(a)--(c) The dynamics of large collective spin states close the equator of a generalized Bloch sphere can be effectively described by first evolving the state in the local tangent plane and then wrapping it back around the sphere. (d)--(f) Exact simulations of the BEC number differences (red histograms) are  in very good agreement with the analytical approximation \eqref{eq:PhaseFlipDistributionApprox} (black lines). The simulation was performed for $N=100$ particles and an initial variance of $\Delta{\sf J}^2_z=N/5$, reached by means of one-axis-squeezing \cite{Kitagawa1993}. The snapshots are taken at times (d) $t_0=0$, (e) $t_1=5.25\pi\hbar/\epsilon$, and (f) $t_2=400 \pi\hbar/\epsilon$ with $\Gamma_{\rm P}=\zeta=0.002\epsilon/\hbar$. At time $t_2$ the distributions have practically converged towards the fully dephased steady state. (g)  In absence of phase diffusion the distribution exhibits (partial) revivals, as illustrated in Panel (g) for time $t_2$. (A complete revival to the state shown in (a) would first be observed at $t=1000\pi\hbar/\epsilon$.)  }
\label{fig:blochspheres}
\end{figure*}

The product of $N$ bosons being in a superposition state (coherent spin state; CSS) can be represented on a generalized Bloch sphere (see Fig.~\ref{fig:blochspheres}), whose polar angle $\theta$ indicates the relative population in $a$ and $b$, while the azimuth $\phi$ is the relative phase of the superposition state. Such a product state $\vert \theta, \phi \rangle$  can be expanded in terms of Dicke states as 
\begin{align}
\left|\theta,\phi\right\rangle \equiv &\frac{1}{\sqrt{(2J)!}}\left(\cos\left(\frac{\theta}{2}\right){\sf c}^{\dagger}_a+e^{i\phi}\sin\left(\frac{\theta}{2}\right){\sf c}^{\dagger}_b\right)^{2J} \vert {\rm vac} \rangle \nonumber\\
=&\sum_{m=-J}^J{2J \choose J+m}\cos\left(\frac{\theta}{2}\right)^{J-m}\sin\left(\frac{\theta}{2}\right)^{J+m}\nonumber\\
&\times e^{-i(J+m)\phi}\left|J,m\right\rangle.
\end{align}
It has minimal and symmetric uncertainties, e.g.\ $\Delta {\sf J}_z^2=\Delta{\sf J}_y^2=|\left\langle{\sf J}_x\right\rangle/2|=J/2$ for $\theta = \pi/2$ and $\phi = 0$.

Applying a nonlinear squeezing operator turns the CSS into a squeezed spin state (SSS) \cite{Kitagawa1993,Ma2011}, which can be useful for metrology \cite{toth2012,toth2014quantum,hosten2016quantum} or robust against dephasing processes \cite{Javanainen1997PhaseDispersion,Schmiedmayer2013}. In addition, it has been demonstrated that the {\em depth of entanglement} increases with squeezing \cite{sorensen2001many,Sorensen2001,Toth2014}, as quantified by the squeezing parameter $\xi^2 = 2(\Delta {\sf J}_{\rm min})^2 / J$.
We note that according to the information-theoretic measure from Ref.~\cite{Froewis2012} 
already the existence of such a state yields a large macroscopicity since  squeezing increases the  quantum Fisher information. 

In terms of the depth of entanglement \cite{Sorensen2001,Toth2014} the non-classicality of SSS lies between a product state (CSS) and the maximally entangled NOON-state $\left|\psi\right\rangle\propto\left|N,0\right\rangle+\left|0,N\right\rangle$, a superposition of all particles being either in mode $a$ or mode $b$. Applying the modification on this NOON state yields a decoherence rate proportional to $N^2$, while that of a product state is proportional to $N$. It  is thus easy to see that a NOON-state with stable phase could serve to exclude a large range of classicalization time scales \cite{bilardello2017collapse}, but they have not been generated experimentally thus far. In contrast, the modification-induced dynamics of SSS, which are frequently realized in experiments, is much more intricate, as discussed in the following.

The free time evolution ${\cal L}\rho = -i [{\sf H},\rho]/\hbar$ of the BEC is characterized by the energy difference $\epsilon$ between the two modes and by the interaction between the particles. Approximating the latter to leading order in ${\sf J}_z$, yields the Hamiltonian \cite{Javanainen1997PhaseDispersion}
\begin{equation}
{\sf H}=\epsilon {\sf J}_z +\hbar\zeta{\sf J}^2_z,
\label{eq:FreeDephasingHamiltonian}
\end{equation}
where $\zeta=d\tilde{\mu}/d (\hbar m)|_{m=0}$ is the change of chemical potential with the occupation difference $m$. Thus, the first term of the Hamiltonian describes rotations around the $z$-axis with angular frequency $\epsilon/\hbar$ on the generalized Bloch sphere, while the second term leads to dispersion.

The experiment starts with the BEC in the state $|\theta=\pi/2,\phi=0\rangle$, which is then squeezed in $z$-direction and  freely evolved for up to 20\,ms. Finally, a $\pi/2$-rotation around the $x$-axis converts the phase distribution into mode occupation differences, which are read-out by time-of-flight measurements, see Fig.\,\ref{fig:2}.

The likelihood required for the hypothesis test is the probability of observing a number difference of $m$ between the two modes,
\begin{align}
P(m | \tau_e,\sigma,I)=& \sum_{J = 0}^{\infty} \left\langle J, m\right|e^{-i\pi{\sf J}_x/2}\rho_t e^{i\pi{\sf J}_x/2}\left| J, m\right\rangle\nonumber\\
= &\sum_{J=0}^\infty P(J|\tau_e,\sigma,I)P_J(m|\tau_e,\sigma,I)\,,
\label{eq:PiHalfPulse}
\end{align} 
where the sum over $J$ accounts for  the possibility of modification-induced particle loss from the BEC during the experiment \cite{Laloe2014}. The modification parameters $\tau_e$ and $\sigma$ enter through the modified time evolution of the state $\rho_t$, which will be discussed next.

\subsection{Double-well potential: phase flips}

Expanding the momentum annihilation operators ${\sf c}({\bf p})$ in Eq.~\eqref{eq:lindblad} in the single-particle eigenmodes in presence of the potential, and neglecting particle loss for the moment (${\mathsf c}_a^\dagger{\mathsf c}_a+{\mathsf c}_b^\dagger{\mathsf c}_b=2J$), yields
\begin{align}
\mathcal{M}_\sigma\rho=&\frac{4 m_{\rm Rb}^2}{\tau_e m_e^2}\int d^3{\bf q}\,f_\sigma(q)\nonumber\\
&\times\left[
{\sf A}({\bf q})\rho{\sf A}^{\dagger}({\bf q})-\frac{1}{2}\{{\sf A}^{\dagger}({\bf q}){\sf A}({\bf q}),\rho\}
\right].
\end{align}   
Here, we used that spatial displacements are negligible on the length scale of the experiment and thus $f_\sigma(q) = \int ds g_\sigma(q,s)$ depends only on $\sigma_q$. The Lindblad operators are given by
\begin{align}
{\sf A}({\bf q})=a_x({\bf q}){\sf J}_x+a_z({\bf q}){\sf J}_z\,,
\end{align}
with
\begin{align}
a_x({\bf q})=&\left\langle \psi_a\right|{\sf W}({\bf q})\left|\psi_b\right\rangle
\nonumber\\
a_z({\bf q})=&i\left\langle \psi_a\right|{\sf W}({\bf q})\left|\psi_a\right\rangle\sin\left(\frac{\Delta_x q_x}{2\hbar}\right)e^{i\Delta_xq_x/2\hbar}.
\label{eq:FlipCoefficients}
\end{align}
Here, $\left|\psi_a\right\rangle$ and $\left|\psi_b\right\rangle$ are the single-atom eigenstates of the two level system with real wavefunctions $\psi_b({\mathbf r})=\psi_a({\mathbf r}-\Delta_x {\mathbf e}_x)\in \mathbb{R}$ and ${\sf W}({\bf q})=\exp(i {\bf q}\cdot \boldsymbol{\mathsf{r}}/\hbar)$
is the momentum transfer operator.

The first part of the Lindblad operator describes rotations around the $x$-axis, or spin-flips, while the second one induces rotations around the $z$-axis, or phase-flips. Such flip operators are frequently used to describe disturbance channels in collective spin states \cite{wang2010sudden,Ma2011}. Since the spatial overlap between the two modes is negligible, $a_x({\bf q})\ll a_z({\bf q})$, the spin-flip contribution will be neglected in the following, implying that $\langle {\sf J}^2_z \rangle_t$ remains constant.

The expectation value of the perpendicular spin components decays as $\left\langle{\sf J}_y\right\rangle_t=e^{-\Gamma_{\rm P} t/2}\left\langle{\sf J}_y\right\rangle_{{\rm f}, t}$ with phase-flip  rate (or dephasing rate)
\begin{align}
\quad\Gamma_{\rm P}=\frac{4 m_{\rm Rb}^2}{\tau_e m_e^2}\int d^3{\bf q}\,f_\sigma(q)|a_z({\bf q})|^2.
\label{eq:SqueezingTEFirstMoments}
\end{align}
Here $\left\langle{\sf J}_y\right\rangle_{{\rm f},t}$ denotes the free time evolution of the  expectation value due to Eq.~\eqref{eq:FreeDephasingHamiltonian}; the same  relation holds for $\langle {\sf J}_x \rangle_t$. Note that the phase-flip decay rate $\Gamma_{\rm P}$ is independent of the degree of squeezing.

The phase-flip operators induce diffusion in the azimuthal plane of the generalized Bloch sphere. The second moment of ${\sf J}_y$ thus evolves as
\begin{align}
\left\langle{\sf J}^2_y\right\rangle_t=\frac{1}{2} \left\langle{\sf J}^2_x + {\sf J}^2_y\right\rangle_{{\rm f},t}
-
\frac{e^{-2\Gamma_{\rm P} t}}{2} \left\langle{\sf J}^2_x - {\sf J}^2_y\right\rangle_{{\rm f},t},
\label{eq:SqueezingTESecondMoments}
\end{align}
and similar for ${\sf J}^2_x$.
For sufficiently large $N$ the squeezing loss rate is  again independent of the initial squeezing since $\left\langle{\sf J}^2_x\right\rangle_{{\rm f},t}\approx J^2$ (as long as oversqueezing is avoided). 

Equations~\eqref{eq:SqueezingTEFirstMoments} and \eqref{eq:SqueezingTESecondMoments}  show that squeezing has no direct implications for the sensitivity on modification-induced decoherence. In contrast to what might be expected intuitively,  an increased {\em depth of entanglement}  does therefore not improve substantially  the macroscopicity of experiments that measure only the first two moments of the collective spin observables.

\subsection{Continuum approximation}

In order to calculate the likelihood \eqref{eq:PiHalfPulse}, we will utilize a continuum approximation on the tangent plane of the generalized Bloch sphere, replacing the discrete probability $P(m|\tau_e,\sigma,I)$ by the continuous probability density $p(m|\tau_e,\sigma,I)$ for real $m$. For this sake, we use that the initial state is aligned with the $x$-axis, $\langle {\sf J}_x \rangle \approx J$, so that
\begin{equation}
 [{\sf J}_y, {\sf J}_z] \approx i J\,,
\end{equation}
which is approximately constant (and not operator valued). Thus we locally replace the sphere by its flat tangent plane and may interpret ${\sf J}_y$ as a position and ${\sf J}_z$ as a momentum operator, see Fig.\,\ref{fig:blochspheres}. The Wigner function of the initial state is then approximated by a Gaussian distribution,
\begin{align} \label{eq:ini}
w_0(j_y,j_z) = &\frac{1}{\sqrt{4\pi^2\left\langle{\sf J}^2_y\right\rangle_0\left\langle{\sf J}^2_z\right\rangle_0}}\nonumber\\
&\times \exp\left[
-\frac{1}{2}\frac{j_y^2}{\left\langle{\sf J}^2_y\right\rangle_0}-\frac{1}{2} \frac{j_z^2}{\left\langle{\sf J}^2_z\right\rangle_0}
\right],
\end{align}
where $(j_y,j_z) \in \mathbb{R}^2$ are continuous variables in the flat tangent plane.

The time evolution of the initial state \eqref{eq:ini} contains the free rotation and dispersion described by Eq.~\eqref{eq:FreeDephasingHamiltonian}, as well as modification-induced dephasing. Representing the dynamics in quantum phase space, the quadratic term in the Hamiltonian \eqref{eq:FreeDephasingHamiltonian} induces shearing in $j_y$, while the linear term leads to a translation in $j_y$ with constant velocity. The phase flips induce diffusion in $j_y$, which increases the variance linearly with time. The corresponding time evolved state can thus be written as
\begin{align} \label{eq:wignert}
 w_t(j_y,j_z) = &\frac{1}{\sqrt{4\pi^2\left(\left\langle{\sf J}^2_y\right\rangle_0+\Gamma_{\rm P} J^2 t\right)\left\langle{\sf J}^2_z\right\rangle_0}}\nonumber\\
&\times \exp\left[
-\frac{1}{2}\frac{(j_y-\epsilon t/\hbar - 2\zeta j_z t)^2}{\left\langle{\sf J}^2_y\right\rangle_0+\Gamma_{\rm P} J^2 t}-\frac{1}{2} \frac{j_z^2}{\left\langle{\sf J}^2_z\right\rangle_0}
\right],
\end{align}
implying that the marginal distribution of $j_z$ remains unaffected by the dynamics.

In order to calculate the likelihood $P_J(m|\tau_e,\sigma,I)=\langle J, m | e^{-i\pi{\sf J}_x/2}\rho_t e^{i\pi{\sf J}_x/2}\left| J, m\right\rangle/P(J|\tau_e,\sigma,I)$ at fixed $J$, we first perform the $\pi/2$-rotation around the $x$-axis, which exchanges $j_y$ and $j_z$ in Eq.~\eqref{eq:wignert}. The resulting distribution is then integrated over $j_y$, and $j_z$ is wrapped back onto the sphere by using $\sin(j_z) = m/J$ and the summation $\int dj_y\sum_k w_t(j_z+2\pi k,j_y)$. This way one obtains the continuous probability density approximating $P_J$,
\begin{widetext}
\begin{align}
p_J(m|\tau_e,\sigma,I)=&\frac{\Theta(J^2-m^2)}{2\pi \sqrt{J^2-m^2}}\left[
\vartheta_3\left(\frac{\arcsin(m/J)-\epsilon t/\hbar}{2},g(t)\right) + \vartheta_3\left(\frac{\pi-\arcsin(m/J)-\epsilon t/\hbar}{2},g(t)\right)
\right],
\label{eq:PhaseFlipDistributionApprox}
\end{align}
\end{widetext}
where
$\Theta(x)$ is the Heaviside function,  $\vartheta_3$ denotes the Jacobi-theta functions of the third kind
\begin{align}
\vartheta_3(u,q)=\sum_{n=-\infty}^{\infty}q^{n^2}e^{2inu}\,,
\end{align}
and the dependence on the initial state is expressed by
\begin{align} \label{eq:gdouble}
g(t)=\exp\left[-\frac{\left\langle{\sf J}^2_y\right\rangle_0}{2J^2}-\frac{\Gamma_{\rm P}t}{2}-2\zeta^2t^2\left\langle{\sf J}^2_z\right\rangle_0\right]\,.
\end{align}

This analytic result captures the generic dephasing effect of random phase flips on a two-mode BEC. The comparison of Eq.~\eqref{eq:PhaseFlipDistributionApprox} with exact numerical calculations shows very good agreement, as demonstrated in Fig.\,\ref{fig:blochspheres}. 

At this stage it might be tempting to use Eq.~\eqref{eq:PhaseFlipDistributionApprox} for Bayesian updating to calculate the macroscopicity. However, since the spatial distance between the two wells of the potential is not much greater than the extension of the modes, the resulting maximizing modification parameters $\sigma$ imply a moderate heating of the BEC. This must be taken into account for a consistent description. A brief discussion of the role of spin flips in single-well potentials will prepare this.

\subsection{Single-well potentials: spin flips}

The dynamics of a BEC in the two lowest eigenstates of a single-well potential, as studied in Ref.~\cite{Schmiedmayer2014}, is strongly affected by spin flips. This marked difference to the double well is due to the spatial overlap between the two modes, see Eq.~\eqref{eq:FlipCoefficients}. The resulting Lindblad operators do not commute with ${\sf J}_z $, but induce additional diffusion in $z$-direction. In combination with the Hamiltonian \eqref{eq:FreeDephasingHamiltonian} this leads to an enhanced dispersion.

If the free rotation frequency $\epsilon/\hbar$ exceeds the spin-flip diffusion rate
\begin{align}
\quad\Gamma_{\rm S}=\frac{4m_{\rm Rb}^2}{\tau_e m_e^2}\int d^3{\bf q}\,f_\sigma(q)|a_x({\bf q})|^2\,,
\end{align}
the average gain in the second moment of $\sf J_z$ can be easily calculated. For times much greater than the rotation period one obtains 
\begin{align}
\left\langle{\sf J}^2_z\right\rangle_t\approx
\frac{\left\langle{\sf J}^2_z\right\rangle_0+J^2}{3}
+\frac{2\left\langle{\sf J}^2_z\right\rangle_0-J^2}{3}
e^{-3\Gamma_{\rm S} t/2}\,.
\label{eq:SpinFlipVarianceBroadening}
\end{align}
For single wells, spin flips will typically dominate the influence of the modification, and phase flips can safely be neglected.

Expanding Eq.~\eqref{eq:SpinFlipVarianceBroadening} for small $\Gamma_{\rm S} t$ and exploiting that $J^2\gg\left\langle {\sf J}^2_z\right\rangle$, yields in the continuum approximation (see App.~\ref{app:a})
\begin{align}
\Delta j_y^2(t)\approx\Delta j_y^2(0)+4\zeta^2 J^2 t^2\left[\left\langle{\sf J}^2_z\right\rangle_0+\frac{\Gamma_{\rm S}J^2 t}{6}\right].
\label{eq:AngleVariance}
\end{align} 
Thus the random spin flips enhance dispersion so that the variance of $j_y$ increases with $t^3$. This results in the probability distribution \eqref{eq:PhaseFlipDistributionApprox} with 
\begin{align}
g(t)=\exp\left[-\frac{\left\langle{\sf J}^2_y\right\rangle_0}{2J^2}-2\zeta^2t^2
\left(\left\langle{\sf J}^2_z\right\rangle_0
+\frac{\Gamma_{\rm S}J^2t}{6}
\right)\right].
\label{eq:SpinFlipGaussApprox}
\end{align} 

In single-well BEC interferometers the modification thus strongly influences the final occupation difference, rendering them attractive for future superposition tests. As explained next, diffusion in the orthogonal $z$-direction is also caused by modification-induced particle loss. The above results can be directly transferred.
 
\subsection{Heating-induced particle loss}
 
In order to include modification-induced particle loss from the BEC, we assume that atoms leaving the two ground modes will never return. This assumption is well justified for a large modification parameter $\sigma_q$, where the particles have a negligible probability of being scattered back to the two lowest modes. 

In this simplified scenario their populations decay exponentially,
\begin{align}
\left\langle{\sf c}^{\dagger}_a{\sf c}_a\right\rangle_t
=e^{-\Gamma_a t}
\left\langle{\sf c}^{\dagger}_a{\sf c}_a\right\rangle_0,\quad
\left\langle{\sf c}^{\dagger}_b{\sf c}_b\right\rangle_t
=e^{-\Gamma_b t}
\left\langle{\sf c}^{\dagger}_b{\sf c}_b\right\rangle_0,
\label{eq:GSpopulation}
\end{align}
with loss rates 
\begin{align}
\Gamma_{a,b}=&\frac{m_{\rm Rb}^2}{\tau_e m_e^2}\int d^3{\bf q}\,f_\sigma(q)\left[1-\left|\left\langle \psi_{a,b}\right|{\sf W}({\bf q})\left|\psi_{a,b}\right\rangle\right|^2\right].
\end{align}
The radius of the generalized Bloch sphere thus decreases with time, and for $\Gamma_a \neq \Gamma_b$ the state is shifted towards one of the poles.

Also the coherences decay exponentially,
\begin{align}
\left\langle{\sf c}^{\dagger}_a{\sf c}_b\right\rangle_t
=e^{-\Gamma_{\rm C} t}
\left\langle{\sf c}^{\dagger}_a{\sf c}_b\right\rangle_0,\quad
\langle{\sf c}^{\dagger}_b{\sf c}_a\rangle_t
=e^{-\Gamma_{\rm C} t}
\langle{\sf c}^{\dagger}_b{\sf c}_a\rangle_0\,,
\label{eq:Coherences}
\end{align}
with 
\begin{align}
\Gamma_{\rm C}=\frac{m_{\rm Rb}^2}{\tau_e m_e^2}\int d^3 {\bf q}\, f_\sigma(q)
\left[
1-\left\langle \psi_a\right|{\sf W}({\bf q})\left|\psi_a\right\rangle
\left\langle \psi_b\right|{\sf W}^{\dagger}({\bf q})\left|\psi_b\right\rangle
\right]\,.
\label{eq:CoherenceDecay}
\end{align}

In order to evaluate the effect of particle loss on the likelihood (\ref{eq:PhaseFlipDistributionApprox}) we use the result of Ref.~\cite{Ma2011} to determine how the variance of $\sf J_{\bf n}$, i.e.\ the angular momentum component in direction ${\bf n}$, changes due to particle loss. Using $J_0,J\gg 1$ one obtains
\begin{align} \label{eq:rescale}
\frac{\left\langle{\sf J}^2_{\bf n}\right\rangle_{J}}{J^2}\approx \frac{\left\langle{\sf J}^2_{\bf n}\right\rangle_{J_0}}{J_0^2}
+\frac{J_0-J}{2J_0J},
\end{align}
where $J$ ($J_0$) is the current (initial) collective spin after the loss of $2(J_0-J)$ particles, and angular brackets $\left\langle \dots\right\rangle_{J}$ denote expectation values after tracing out the lost particles. The second term shows that the rescaled second moment $\left\langle{\sf J}^2_{\bf n}\right\rangle_{J}/{J^2}$ increases due to the particle loss.

Combining Eq.~\eqref{eq:rescale} with Eq.~(\ref{eq:GSpopulation}), using that in the double-well  $\Gamma_a=\Gamma_b \equiv \Gamma_{\rm L}$, expanding the result to linear order in $\Gamma_{\rm L}t$, and finally repeating the steps carried out in the previous section to account for simultaneous shearing and diffusion, yields the distribution (\ref{eq:PhaseFlipDistributionApprox}) with 
\begin{align}
g(t)=&\exp\left[-\frac{\left\langle{\sf J}^2_y\right\rangle_0}{2J_0^2}
-\frac{\Gamma_{\rm P}t}{2}-\frac{\Gamma_{\rm L}t}{4J_0}
\right.\nonumber\\
&\left.-2\zeta^2t^2
\left(\left\langle{\sf J}^2_z\right\rangle_0
+\frac{\Gamma_{\rm L} J_0 t}{6}
\right)\right].
\label{eq:ContinuousDoubleWellGauss}
\end{align}

The enhancement of the dispersion looks similar to the single-well case \eqref{eq:SpinFlipGaussApprox}, but it is weaker by the (significant) factor $1/J_0$. Note that the dispersion rate $\zeta$ decreases with decreasing $J_0$, and the linear approximation of the chemical potential leading to the free Hamiltonian (\ref{eq:FreeDephasingHamiltonian}) will fail if too many particles are lost.  

The  distribution of the remaining particles turns out to be binomial \cite{schrinski2017sensing} given that $\Gamma_a=\Gamma_b\equiv \Gamma_{\rm L}(\tau_e,\sigma,I)$. The  probability density for $m \in \mathbb{R}$, i.e. the continuous approximation of Eq.~\eqref{eq:PiHalfPulse}, therefore takes the final form
\begin{align}\label{eq:37}
p(m|\tau_e,\sigma,I) = & \sum_{J=0}^{J_0}{{J_0}\choose{J}}\left(1-e^{-\Gamma_{\rm L}t}\right)^{J}\left (e^{-\Gamma_{\rm L}t}\right)^{J_0-J} \nonumber \\
 & \times p_J(m|\tau_e,\sigma,I),
\end{align}
where $p_J(m|\tau_e,\sigma,I)$ is given by Eqs.~\eqref{eq:PhaseFlipDistributionApprox} and \eqref{eq:ContinuousDoubleWellGauss} and $p_0(m|\tau_e,\sigma,I) = \delta(m)$. This equation can now be used for the Bayesian updating procedure \eqref{eq:likelihoodproduct} and for evaluating the macroscopicity \eqref{eq:Macroscopicity}.

\subsection{Experimental parameters}

The BEC reported in Ref.\ \cite{Schmiedmayer2013} consists of $N=2J_0\approx1200$ $^{87}$Rb atoms in a double-well configuration with a spatial separation of $\Delta_x\approx2\,\mu{\rm m}$ in $x$-direction and an initial number squeezing of $\Delta {\mathsf J}_z^2=0.41^2 J_0/2$. The trapping frequencies are $\omega_x/2\pi=1.44\,{\rm kHz}$, $\omega_y/2\pi=1.84\,{\rm kHz}$ and $\omega_z/2\pi=13.2\,{\rm Hz}$, so that the motion in $z$-direction is quasi-free. The two lowest energy levels of this potential have a gap of $\epsilon/\hbar=2.19\,{\rm kHz}$ and the first order corrections of the chemical potential are characterized by
$\zeta=4\,$Hz.

Approximating the ground states harmonically with the  widths $\sigma_{x,y}=\sqrt{\hbar/2m_{\rm Rb}\omega_{x,y}}$ yields the phase-flip and loss rates
\begin{align}
\Gamma_{\rm P}= & \frac{2m_{\rm Rb}^2}{\tau_e m_e^2} \frac{1-\exp[-\Delta_x^2\sigma_q^2/(4\sigma_q^2\sigma_x^2+2\hbar^2)]}{\sqrt{(1+2\sigma_q^2\sigma_x^2/\hbar^2)(1+2\sigma_q^2\sigma_y^2/\hbar^2)}} \\ 
\Gamma_{\rm L}= &\frac{m_{\rm Rb}^2}{\tau_e m_e^2}\left(1-\frac{1}{\sqrt{(1+2\sigma_q^2\sigma_x^2/\hbar^2)(1+2\sigma_q^2\sigma_y^2/\hbar^2)}}\right).
\label{eq:Gl}
\end{align}

For the experimental parameters given above, the particle loss rate $\Gamma_{\rm L}$ cannot be neglected compared to the phase-flip rate $\Gamma_{\rm P}$ in the entire parameter regime of $\sigma$. This is due to the fact that the widths of the ground state modes $\sigma_{x,y}$ are comparable to the spatial separation of the wells $\Delta_x$. Consequently, it cannot be excluded that the observed lack of particle loss due to modification-induced heating may significantly affect the hypothesis test, even though confirming the conservation of particle number does not verify quantum coherence. 

As a remedy, we condition the likelihood \eqref{eq:37} on the observed particle number, as explained at the end of  Sect.~\ref{sec:2A}. This makes the overall atom number part of the experimental background information, and we can separately assess the modification-induced loss of interference visibility \emph{given that} a certain particle number was detected. The conditioned likelihood \eqref{eq:10cl} is obtained by dividing the likelihood  \eqref{eq:37} by the probability
\begin{align}
P(d_{\rm heat}|\tau_e,\sigma, I) = & \sum_{J=\lfloor 0.9J_0\rfloor}^{J_0}{{J_0}\choose{J}}\left(e^{\Gamma_{\rm L}t}-1\right)^{J}e^{-J_0\Gamma_{\rm L}t}\,
\end{align}
that not more than 10\% of the particles are lost, 
$d_{\rm heat}:=\{J\geq 0.9\,J_0\}$. 
This threshold value is taken as a conservative estimate given that the number of the trapped particles fluctuates by at most 10\% between the individual experimental runs.

All information is now available to perform the Bayesian hypothesis test, as described in Sect.~\ref{sec:2} using the 1438 data points presented in Fig.~\ref{fig:2}(b). Numerical maximization of $\tau_{\rm m}(\sigma)$ yields a macroscopicity value of $\mu_{\rm m}=8.5$.
The maximum of $\tau_{\rm m}(\sigma)$  is attained for the modification parameter $\sigma_q\simeq\hbar/0.77\,{\rm mm}$. As one would expect, this 
roughly corresponds to the parameter value  where the phase-flip rate is maximized (at $\Gamma_{\rm P}=1.7/\tau_e$), implying that dephasing is most pronounced. The corresponding particle loss rate is an order of magnitude lower ($\Gamma_{\rm L}=0.11/\tau_e$).

The macroscopicity attained in the double-well BEC interferometer is comparable to the value expected for an atom interferometer operating  single Rubidium atoms on the same timescale. For instance, using the estimate in \cite{Nimmrichter2013} with an interference visibility $f=0.2$ after $t=20\,$ms, one would also obtain $\mu=8.5$. This close match might be expected for an unsqueezed BEC, where all atoms are uncorrelated. That the number squeezed BEC discussed here does not reach an appreciably higher macroscopicity, despite its large depth of entanglement, can be attributed to the fact that single-particle observables are measured. They are not sensitive to many-particle correlations that are potentially destroyed by the classicalizing modification.
In contrast, if the modification had induced spin flips, as in a single-well interferometer scenario \cite{Schmiedmayer2014},  the resulting destruction of  number-squeezing could be observed 
due to the interplay between the modification effect and the intrinsic dispersion caused by atom-atom interactions,
see Eq.~\eqref{eq:SpinFlipGaussApprox}.

\section{Leggett-Garg test with an atomic quantum random walk} \label{sec:4}

\subsection{Setup}

Reference~\cite{Robens2015} describes a test of the Leggett-Garg inequality with single atoms performing a quantum random walk in an optical lattice formed by two circularly polarized laser beams. The form of the lattice potential depends on the hyperfine state of the atoms, so that by preparing single $^{133}$Cs atoms in a superposition of two hyperfine states and displacing the two lattices in opposite directions, one can prepare the atom in a superposition  of left- and right-directed movements. We denote the  displacement length of a single step by $d$, and the associated time required to displace the lattices by $T_{\rm d}$.

The quantum random walk (Fig.\,\ref{fig:3}) is performed by first applying a $\pi/2$-pulse over the duration $T_{\rm r}$, which prepares the atom in a superposition of the hyperfine states and then transforming this into a spatial superposition by displacing the lattices for the duration $T_{\rm d}$. This scheme is iterated four times and finally a position measurement of the atom is performed, collapsing its position into a definite lattice site. Since no $\pi/2$-pulse is applied after the fourth step, atoms which do not end up in the same hyperfine state are excluded by the measurement protocol. This means that all paths which contribute to the interference must recombine after the third step.

Representing the two-level internal degree of freedom by a spinor, the action of a single step in the quantum random walk is given by the unitary operator
\begin{align} \label{eq:trafo}
{\sf S}=\frac{1}{\sqrt{2}}
\begin{pmatrix}
{\sf U}_d & -{\sf U}_d \\
{\sf U}_{d}^\dagger & {\sf U}_{d}^\dagger 
\end{pmatrix},
\end{align}
with the translation operator ${\sf U}_d=\exp (-i{\sf p}d/2\hbar )$. A straight-forward calculation shows that in addition to the classical random-walk trajectories, involving no coherences, there are only two classes of trajectories contributing to the interference pattern, see Fig. \ref{fig:QRWSkizze}: (i) the atomic wavefunction is split and recombines immediately in the following step; (ii) the atomic wavefunction is split in the first step, then both parts are displaced either to the left or the right in the second step, and they recombine in the third step. To model the experimental outcome, one has to determine the likelihood
\begin{equation} \label{eq:qrwlike}
 P(\ell | \tau_e,\sigma,I) = {\rm tr}_{\rm spin} \left(\langle \ell | \rho | \ell \rangle\right ),
\end{equation}
where $\ell\in \{-2,-1,0,1,2\}$ labels the lattice sites that can be reached in four steps and $\rho$ is the final state evolved  under influence of the modification \eqref{eq:MIMpointparticle} with parameters $\tau_e$ and $\sigma$.

\subsection{Impact of the modification}

\begin{figure}
  \centering
  \includegraphics[width=0.45\textwidth]{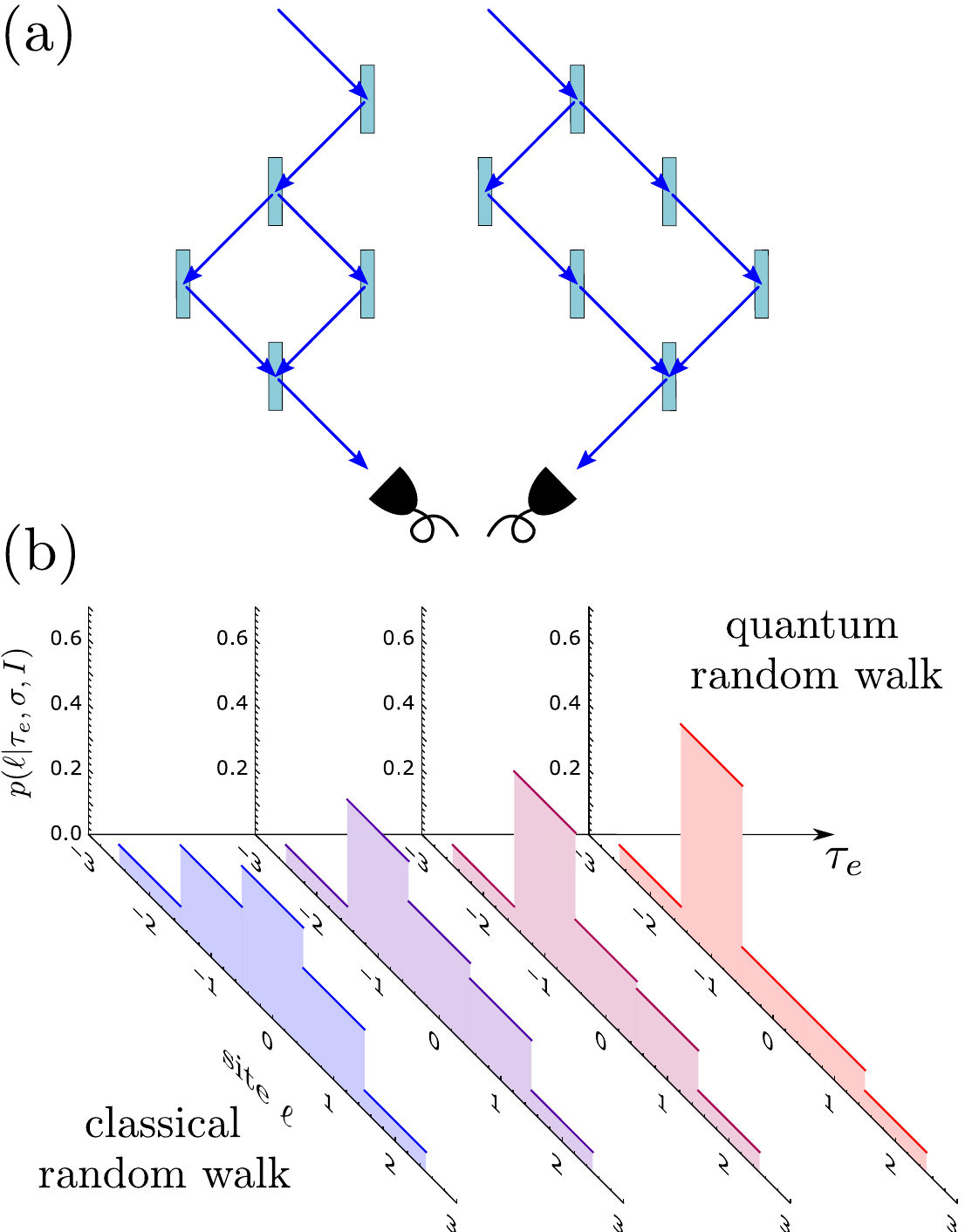}
  \caption{(a) Examples of the two classes of coherently split trajectories contributing to the quantum random walk: (i) the atomic wavefunction splits in the first or second step and recombines afterwards; (ii) the atomic wavefunction splits in the first step, then both parts move one step in parallel, and recombine in the third step. (b) Quantum-to-classical transition of the quantum random walk with decreasing classicalization timescale $\tau_e$. The diagrams depict the final-site probabilities \eqref{eq:QRW4} for modification parameters $\hbar/\sigma_q=d/10$ and  $\tau_e m_e^2/m_{\rm Cs}^2=1\,\mu{\rm s},50\,\mu{\rm s},100\,\mu{\rm s},10\,{\rm ms}$ from left to right.}
\label{fig:QRWSkizze}
\end{figure}

Since the separation between neighboring lattice sides is $d = 433$\,nm, spatial displacements can be neglected in the modification \eqref{eq:MIMpointparticle}, i.e.\ we can set $\sigma_s =0$. The influence of the modification on a superposition of momentum states can be calculated by drawing on the results in Ref.~\cite{schrinski2017sensing}, where the momentum superposition of a non-interacting BEC in the limit of a high number of atoms was approximated by a macroscopic wave function (obeying the single particle Schr\"odinger equation). One can directly carry over these results to the present case of a single Cesium atom. As a result, the likelihood \eqref{eq:qrwlike} can be calculated with the help of the dimensionless coherence reduction factor
\begin{align} \label{eq:redfac}
 R(t) = & \exp\left[
-\frac{2 T_{\rm d}m_{\rm Cs}^2}{\tau_e m_e^2} \left (1  - \frac{\sqrt{\pi} \hbar}{\sqrt{2} d \sigma_q}\erf \left (\frac{d \sigma_q}{\sqrt{2} \hbar } \right ) \right ) \right ] \nonumber \\
& \times \exp\left[
-\frac{t m_{\rm Cs}^2}{\tau_e m_e^2} \left (1
- \exp \left ( -\frac{d^2 \sigma_q^2}{2 \hbar^2} \right ) \right ) \right],
\end{align}
where $t$ is the time over which the superposition state is maintained at a constant distance of $d$. Thus, in the case of the path (i) $t = T_{\rm r}$, and in case of path (ii) $t = T_{\rm d} + 2 T_{\rm r}$.

Initializing the random walk in the upper hyperfine state,
one can identify all contributing trajectories by applying Eq.~\eqref{eq:trafo} four times. After weighting these with the appropriate  reduction factors \eqref{eq:redfac},  the trace \eqref{eq:qrwlike} finally yields the probability distribution\footnote{Starting with the lower hyperfine state one obtains the mirrored version of the distribution \eqref{eq:QRW4}.} 
\begin{subequations}
\label{eq:QRW4}
\begin{align} 
P(-2|\tau_e,\sigma,I)=&\frac{1}{16}, \\
P(-1|\tau_e,\sigma,I)= & \frac{1}{4} + \frac{1}{4}R(T_{\rm r})+\frac{1}{8}R(T_{\rm d} + 2T_{\rm r})\\
P(0|\tau_e,\sigma,I)=&\frac{3}{8}- \frac{1}{4}R(T_{\rm r}),\\
P(1|\tau_e,\sigma,I)=& \frac{1}{4}- \frac{1}{8}R(T_{\rm d} + 2T_{\rm r}),\\
P(2|\tau_e,\sigma,I)= & \frac{1}{16}.
\end{align}   
\end{subequations}
These results reflect what is to be expected from a classicalizing modification applied to the quantum random walk: The classical random walk probabilities are retrieved in the limit $\tau_e \to 0$, where $R(t) = 0$, while the opposite limit $\tau_e \to \infty$, i.e.\ $R(t) = 1$, yields the
ideal quantum random walk probabilities. The gradual transition between classical and quantum behavior is depicted in Fig.~\ref{fig:QRWSkizze}.

In the Leggett-Garg test of Ref.~\cite{Robens2015} additional measurement results were postselected conditioned on whether the walker moves in the first step  to the left or to the right. In this case the random walk effectively starts one step later, and thus only trajectories of type (i) contribute to the interference. The resulting probabilities can be determined as above,
\begin{subequations}
\label{eq:QRW3}
\begin{eqnarray}
P_{\rm L}(-2|\tau_e,\sigma,I)& = & P_{\rm R}(2|\tau_e,\sigma,I)=\frac{1}{8}, \\
P_{\rm L}(-1|\tau_e,\sigma,I)& = & P_{\rm R}(1|\tau_e,\sigma,I)=  \frac{3}{8}+ \frac{1}{4} R(T_{\rm r}), \\
P_{\rm L}(0|\tau_e,\sigma,I)&= & P_{\rm R}(0|\tau_e,\sigma,I)= \frac{3}{8} - \frac{1}{4} R(T_{\rm r}), \\
P_{\rm L}(1|\tau_e,\sigma,I)&= &P_{\rm R}(-1|\tau_e,\sigma,I)= \frac{1}{8},\\
P_{\rm L}(2|\tau_e,\sigma,I) &= &P_{\rm R}(-2|\tau_e,\sigma,I)=0.
\end{eqnarray} 
\end{subequations}
The subscripts L or R denote that the first step was performed to the left or right.

For completeness, we note that the Leggett-Garg inequality studied in \cite{Robens2015} reads as
\begin{align}
\sum_{\ell=-2}^2 {\rm sgn}(\ell)\left (P(\ell) - \frac{1}{2} \left [ P_{\rm L}(\ell)+P_{\rm R}(\ell) \right ] \right )
\leq 0,
\end{align} 
where we dropped the parameters $\tau_e,\sigma,I$ for brevity. This Leggett-Garg inequality can be rewritten in terms of the modification parameters through the reduction factor \eqref{eq:redfac} by inserting Eqs. (\ref{eq:QRW4}) and (\ref{eq:QRW3}), 
\begin{align}
R(T_{\rm r}) + R(T_{\rm d} + 2 T_{\rm r}) \le 0.
\label{eq:LGTInequalityMIM}
\end{align}
This inequality is always violated unless $\tau_e $ vanishes, but the left-hand side approaches zero exponentially with decreasing $\tau_e$. Note that our assessment of macroscopicity is not based on such a derived quantity, but on the raw data of detection clicks.

\subsection{Experimental parameters}

In the experiment the displacement and resting time are $T_{\rm d}=21\,\mu{\rm s}$ and $T_{\rm r}=5\,\mu{\rm s}$ and the distance between each lattice site is $d=433\,{\rm nm}$. Maximizing the effect of the modification we note that the reduction factor \eqref{eq:redfac} decreases with increasing $\sigma_q$ and that the five percent quantile $\tau_{\rm m}(\sigma)$ saturates for $\hbar/\sigma_q \ll d$. To assess  the macroscopicity, we take the value $\hbar/\sigma_q \approx d/10$, where $\tau_{\rm m}(\sigma)$ already takes the saturated value, yielding $\mu_{\rm m}=7.1$.

Finally, since we neglected possible effects of modification-induced heating so far, we have to verify that this is justified here, i.e. at the stated value of $\sigma_q$ and for the relevant range of classicalization time scales $\tau_{e }$. This can be done conservatively by calculating the heating rate with the 5\% quantile of Jeffreys' prior ($\tau_{e} \simeq 10^6\,{\rm s}$). It serves as an upper bound (see Fig.~\ref{fig:3}) due to Bayesian updating. The resulting  temperature increase of $\Delta T\approx 6\,\mu{\rm K}$ over the duration of the whole experiment is moderate, amounting to less than 1/13 of the potential depth. It thus renders particle loss negligible, so that no explicit conditioning on a likelihood which accounts for heating is required to arrive at \eqref{eq:QRW4} and \eqref{eq:QRW3}.

In summary, the macroscopicity of the  atomic Leggett-Garg test is dominated by the timescale on which the experiment was performed, i.e.\ the ramp- and waiting-time between random walk steps. Since only neighboring trajectories contribute to interference, the relevant length scale of the superposition state is given by the lattice spacing $d$ rather than by the spatial extension of the final state. This could be enhanced by implementing  a $\pi/2$-pulse after the fourth step, or by performing more steps, so that  also trajectories separated by more distant sites contribute to the interference pattern.

\section{Mechanical entanglement of photonic crystals} \label{sec:5}

\subsection{Measurement protocol}

The observation of entanglement  between two nanomechanical oscillators reported in Ref.~\cite{Riedinger2018} is based on a coincidence measurement of Stokes- and anti-Stokes photons created in photonic crystal nanobeams placed in the two arms of a Mach-Zehnder interferometer, see Fig.~\ref{fig:4}.  In the first step (pump), a photon is sent through the entrance beam splitter, excites a single phonon in one of the two nanobeams, thereby creating entanglement in their mechanical excitation. The Stokes-scattered photon is detected behind the exit beam splitter. In the second step (read), a further photon enters the interferometer through the entrance beam splitter, leading to stimulated emission in the photonic crystal. The resulting anti-Stokes scattered photon, which serves to read out the entanglement, is also detected behind the exit beam splitter.

We denote the measurement outcomes of the Stokes and the anti-Stokes photon detectors by $\pm_{1,2}$, where $+$ ($-$) refers to the upper (lower) detector behind the exit beam splitter and the index refers to the pump and read photon, respectively. The likelihood for a certain coincidence measurement is
\begin{equation} \label{eq:likenano}
 P(\pm_1,\pm_2|\tau_e,\sigma,I) =  {\rm tr} \left (| \pm_1, \pm_2 \rangle\langle \pm_1, \pm_2| \rho_{\rm fin} \right ) 
\end{equation}
where $\rho_{\rm fin}$ is the total final state of both oscillators and both photons. The modification parameters $\tau_e$ and $\sigma$ only enter through their influence on the dynamics of the nanomechanical oscillators.

In each nanobeam a single mechanical mode contributes to the measurement signal of the experiment. Even though the pump photon can excite this mode only once, we will in the following allow for arbitrary phonon occupations $|k, \ell \rangle$ of the two oscillators to account for modification-induced heating.

Given that the two relevant oscillator modes are initially in the ground state, the total wave function of the system after the pump photon traversed the exit beam splitter reads
\begin{align}
\left|\psi\right\rangle_{t=0} =&
\frac{1}{2}\left[
\left|+\right\rangle_1\left(\left|1,0\right\rangle+e^{i\phi}\left|0,1\right\rangle\right)\right.\nonumber\\
&\left.+\left|-\right\rangle_1\left(\left|1,0\right\rangle-e^{i\phi}\left|0,1\right\rangle\right)
\right] | {\rm vac} \rangle_2,
\label{eq:MHOInitialStateSP} 
\end{align}
where $\phi$ is the initial relative phase. The state \eqref{eq:MHOInitialStateSP} now evolves freely according to the modified master equation \eqref{eq:modvonneum} into the mixed state $\rho_t$ until the read photon passes the interferometer.

The measurement with the read photon can be described through application of the read operator ${\sf R}$, as $\rho_{\rm fin}={\sf R} \rho_t {\sf R}^\dagger/\mathcal{N}$. Here, the factor $\mathcal{N}={\rm tr}({\sf R}^\dagger{\sf R} \rho_t )$ accounts for the conditioning on coincident detections of Stokes and  anti-Stokes photons. The read operator ${\sf R}$ first annihilates a phonon in one of the two oscillators and simultaneously creates a read photon in the corresponding interferometer arm,
with the relative phase $\theta$  between the two arms determined by the experimental setup. In a second step, the thus created photon traverses again the beam splitter, yielding in total 
\begin{align} \label{eq:read}
 {\sf R} |\pm\rangle_1|k, \ell \rangle |{\rm vac}\rangle_2 = & \frac{|\pm\rangle_1}{\sqrt{2k+2\ell}} \left [\sqrt{k} |k-1,\ell \rangle \left ( |+\rangle_2 + |-\rangle_2 \right ) \right. \nonumber \\
  & \left. + e^{i \theta} \sqrt{\ell} |k,\ell-1\rangle  \left ( |+\rangle_2 - |-\rangle_2 \right ) \right ]
\end{align}
for $(k,\ell) \neq (0,0)$. By in addition setting ${\sf R}|\pm\rangle_1 |0, 0 \rangle |{\rm vac}\rangle_2=0 $ we account for the fact that the phonon ground state (which  may be populated by modification-induced transitions)  cannot lead to a coincidence detection involving an anti-Stokes photon. 

The probability \eqref{eq:likenano} can be written as due to a generalized measurement,  $P(\pm_1,\pm_2|\tau_e,\sigma,I) = {\rm tr} ({\sf F}_{\pm_2} \rho^{(\pm_1)}_t )/\mathcal{N}$. Here, the oscillator state 
\begin{align}\label{eq:rhopm}
\rho^{(\pm_1)}_t=\langle \pm|_1\langle{\rm vac}|_2\rho_t|{\rm vac}\rangle_2| \pm\rangle_1
\end{align}
 is conditioned on the detection of the Stokes photon, and ${\sf F}_{\pm_2} = {\rm tr}_1 (\langle {\rm vac} |_2 {\sf R}^\dagger | \pm \rangle_2 \langle \pm |_2 {\sf R} | {\rm vac} \rangle_2)$ describes the measurement of the anti-Stokes photon, 
\begin{align}
{\sf F}_{\pm_2} = & \frac{1}{2}\left(\sum_{k=1,\ell=0}^{\infty}\frac{k}{k+\ell}\left|k,\ell\right\rangle\left\langle k,\ell\right|+\sum_{k=0,\ell=1}^{\infty}\frac{\ell}{k+\ell}\left|k,\ell\right\rangle\left\langle k,\ell\right|\right.\nonumber \\
&\left.\pm\sum_{k=1,\ell=0}^{\infty}e^{i\theta}\frac{\sqrt{k(\ell+1)}}{k+\ell}\left|k,\ell\right\rangle\left\langle k-1,\ell+1\right|\right.\nonumber\\
&\left.\pm\sum_{k=0,\ell=1}^{\infty}e^{-i\theta}\frac{\sqrt{(k+1)\ell}}{k+\ell}\left|k,\ell\right\rangle\left\langle k+1,\ell-1\right|\right) \,.
\label{eq:MHORiedingerProjectorPositive}
\end{align}

To prepare the calculation of the likelihoods, we  now determine the influence of the modification on the initial oscillator state \eqref{eq:rhopm}.

\subsection{Impact of the modification}

To handle the elastic deformation of a single nanomechanical beam, we
first note that all atoms in the solid can be treated as distinguishable. One can therefore use the Lindblad operators \eqref{eq:lindblad} in first quantization, 
\begin{align}\label{eq:lindblad2}
{\sf L}({\bf q},{\bf s})
=&\sum_n\frac{m_n}{m_e}\exp\left[-i\frac{\textbf{\textsf{r}}_n\cdot{\bf q}-\textbf{\textsf{p}}_n\cdot{\bf s}}{\hbar}\right]\,.
\end{align} 
To express this in terms of the mode variables, we  expand the position operator $\textbf{\textsf{r}}_n$ of each individual atom around its equilibrium position ${\bf r}_n^{(0)}$,
\begin{align}
\textbf{\textsf{r}}_n={\bf r}_n^{(0)}+{\bf w}({\bf r}_n^{(0)}){\sf Q}\,,
\end{align} 
in terms of the classical mode function \cite{madelung2012introduction,fetter2003theoretical} of the relevant displacement mode  ${\bf w}({\bf r})$   and its operator-valued amplitude ${\sf Q}$. The latter can also be written using the mode creation and annihilation operators ${\sf a}^\dagger$ and ${\sf a}$,
\begin{align} \label{eq:displfield}
{\sf Q}
=\sqrt{\frac{\hbar}{2\varrho V_{\rm m}\omega}}\left({\sf a}+{\sf a}^{\dagger}\right),
\end{align} 
where $\varrho$ is the mass density of the material, $\omega$ the mechanical frequency, and $V_{\rm m}$ the mode volume, see App.\ \ref{app:b}. 

Accordingly,  the momentum operator in \eqref{eq:lindblad2} takes the form
\begin{align} \label{eq:momfield}
\textbf{\textsf{p}}_n =\frac{m_n}{\varrho V_{\rm m}}{\bf w}({\bf r}^{(0)}_n){\sf P}= i \sqrt{\frac{\hbar \omega_k m_n^2}{2\varrho V_{\rm m}}}{\bf w}({\bf r}^{(0)}_n)\left({\sf a}^{\dagger}-{\sf a}\right).
\end{align} 
This equation implies that the modification-induced spatial displacement $\mathbf{s}$ in \eqref{eq:lindblad2} scales with the mass of the atom divided by the effective mass of the mechanical mode, which is on the order of the nanobeam mass. The spatial displacement is therefore negligible for all scenarios that lead to observable decoherence, 
allowing us to approximate the Lindblad operators as
\begin{align}
&{\sf L}({\bf q})
\simeq\sum_n\frac{m_n}{m_e}\exp\left[-\frac{i}{\hbar}\left({\bf r}_n^{(0)}+\sum_k{\bf w}_k({\bf r}^{(0)}_n){\sf Q}_k\right)\cdot{\bf q}\right]
\nonumber\\
&= \frac{1}{m_e}\int d^3{\bf r}\,\varrho({\bf r}) \exp\left[-\frac{i}{\hbar}\left({\bf r}+\sum_k{\bf w}_k({\bf r}){\sf Q}_k\right)\cdot{\bf q}\right],
\label{eq:HarmOscFullLindbladOp}
\end{align}
where $k$ is a mode index, and $\varrho({\bf r})=\sum_n m_n\delta({\bf r}-{\bf r}^{(0)}_n)$ denotes the mass density of the oscillator. The latter can be replaced by a continuous, homogeneous mass density provided the characteristic length scale  $\hbar/\sigma_q$ is much greater than the lattice spacing of the crystal structure.

The Lindblad operators \eqref{eq:HarmOscFullLindbladOp} may be expanded to first order in the relevant mode amplitude ${\sf Q}$ as long as $\sigma_q \ll \sqrt{2 \varrho V_{\rm m} \omega\hbar}$. This decouples the different modes and we have
\begin{align}
{\sf L}({\bf q})=-\frac{i}{\hbar}\left[\widetilde{{\bf w}}_{\varrho}({\bf q})\cdot{\bf q}\right]{\sf Q}, 
\label{eq:HarmOscDiffLO}
\end{align}  
where we introduced
\begin{align}
\widetilde{{\bf w}}_{\varrho}({\bf q})&=
\frac{1}{m_e}\int d^3{\bf r}\,\varrho({\bf r}){\bf w}({\bf r})e^{-i{\bf r}\cdot{\bf q}/\hbar}.
\end{align}

The total master equation including the free harmonic Hamiltonian and the Lindblad operators \eqref{eq:HarmOscDiffLO} of both oscillators can be solved analytically with the help of the characteristic function
\begin{align}
\chi_t({\bf Q},{\bf P})=\int d^2{\bf Q}'\,e^{i{\bf P}\cdot{\bf Q}'/\hbar}\left\langle {\bf Q}'+\frac{{\bf Q}}{2}\right|\rho_t\left|{\bf Q}'-\frac{{\bf Q}}{2}\right\rangle,
\end{align}
where ${\bf Q} = (Q_1,Q_2)$ and ${\bf P} = (P_1,P_2)$ are the joint  position and momentum coordinates of both oscillators. The evolution equation for the characteristic function reads
\begin{align}
\partial_t\chi_t({\bf Q},{\bf P})=&
\left(- \frac{1}{\varrho V_{\rm m}}{\bf P}\cdot \nabla_{\bf Q}+\varrho V_{\rm m}{\bf Q}\cdot{\rm \Omega}^2\nabla_{\bf P}-\frac{U(\sigma) {\bf Q}^2}{\tau_e}\right)\nonumber\\
&\times\chi_t({\bf Q},{\bf P}),
\label{eq:HarmOscDiffCharTE}
\end{align} 
where ${\rm \Omega}$ is the diagonal matrix containing the two slightly detuned frequencies of both oscillators and
\begin{align}
U(\sigma)=&\frac{1}{2\hbar^2}\int d^3 {\bf q}\, f_\sigma(q) \left|\widetilde{{\bf w}}_{\varrho}({\bf q})\cdot {\bf q}\right|^2 .
\label{eq:HarmOscDiffFT}
\end{align}
Here we exploited that the separation of the two oscillators is much greater than $\hbar/\sigma_q$.

The time evolved characteristic function is given by
\begin{align} \label{eq:soldifflim}
\chi_t({\bf Q},{\bf P})=&
\exp\left[
-\frac{U(\sigma)}{\tau_e}\int_0^t dt' {\bf Q}^2_{t'} \right]\chi_0({\bf Q}_t,{\bf P}_t),
\end{align}
with
\begin{align}
{\bf Q}_t=&\cos({\rm \Omega t}){\bf Q}+\frac{1}{\varrho V_{\rm m}}{\rm \Omega}^{-1}\sin({\rm \Omega t}){\bf P}\nonumber\\
{\bf P}_t=&\cos({\rm \Omega t}){\bf P}-\varrho V_{\rm m}{\rm \Omega}\sin({\rm \Omega t}){\bf Q}.
\label{eq:MarmoOscFreeQP}
\end{align}

Calculating the initial characteristic function of the state \eqref{eq:rhopm}  and evaluating \eqref{eq:HarmOscDiffFT} for a given mode function ${\bf w}({\bf r})$ allows one to determine analytically  the likelihoods \eqref{eq:likenano}. 

\subsection{Particle loss}

For increasing $\sigma_q$ the energy gain induced by momentum translations due to the Lindblad operators \eqref{eq:lindblad2} can exceed the binding energy of the silicon atoms in the crystal. Thus, the modification may induce particle loss already deep in the diffusive regime. The solution \eqref{eq:soldifflim} of the mode dynamics cannot capture this because the mode expansion assumes the atoms to reside in infinitely extended harmonic potentials. Due to the finiteness of the real binding potential there is a critical momentum transfer $q_c$ beyond which the sole effect of the modification is a reduction of the atom number in the crystal. 

To account for this particle loss, we split Eq.~\eqref{eq:MIMpointparticle} into a part $\mathcal{M}_\sigma^{<}$ with  momentum transfers $|{\bf q}|<q_c$ that will most likely leave the atoms in the crystal, and into the part $\mathcal{M}_\sigma^{>}$ with $|{\bf q}|>q_c$ removing them into the vacuum,
\begin{align}\label{eq:LindbladSum}
\mathcal{M}_\sigma\rho_t&=\int\displaylimits_{q<q_c} d^3 {\bf q}\,f_\sigma(q)\left[{\sf L}({\mathbf q})\rho {\sf L}^\dagger({\mathbf q})-\frac{1}{2}\left\{{\sf L}^\dagger({\mathbf q}){\sf L}({\mathbf q}),\rho\right\}\right]\nonumber\\
&+\int\displaylimits_{q>q_c} d^3 {\bf q}\,f_\sigma(q)\left[{\sf L}({\mathbf q})\rho {\sf L}^\dagger({\mathbf q})-\frac{1}{2}\left\{{\sf L}^\dagger({\mathbf q}){\sf L}({\mathbf q}),\rho\right\}\right].
\end{align} 
A Dyson expansion shows that the final state can be written as a sum
\begin{align}
\rho_t=\exp\left[\frac{t}{i\hbar}\mathcal{H}+\frac{t}{\tau_e}\mathcal{M}_\sigma^{<}\right]\rho_0+\widetilde{\rho}
\end{align}
where only the first term is consistent with the coincidence measurement \eqref{eq:likenano}.
Its reduced trace can be absorbed in the normalization $\mathcal{N}$ reflecting the conditioning on the coincidence measurements.

The time evolution under the modification $\mathcal{M}_\sigma^{<}/{\tau_e}$ can now be treated as in the previous section, yielding Eq.~\eqref{eq:soldifflim} with 
$U(\sigma)$ replaced by 
\begin{align}\label{eq:HarmOscDiffFTParticleLoss}
U_{<}(\sigma)=&\frac{1}{2\hbar^2}\int_{q<q_c} d^3 {\bf q}\, f_\sigma(q) \left|\widetilde{{\bf w}}_{\varrho}({\bf q})\cdot {\bf q}\right|^2 .
\end{align}

\subsection{Experimentally achieved macroscopicity}

The two oscillators in Ref.~\cite{Riedinger2018} are characterized by the effective mass $\varrho V_{\rm m} \approx 9\times10^{-17}$\,kg \cite{PrivateCommRiedinger} and the mechanical frequency $\omega \approx 2\pi\times 5\,{\rm GHz}$. The exact displacement field depends on the precise geometry of the photonic crystal, and is only numerically accessible. Since the details of the mode function are expected to be of minor relevance we approximate the shape of the oscillator by an elastic silicon cuboid containing only those atoms of the nanobeam that contribute to the elastic deformation. The resulting displacement field of the simplest longitudinal mode has the form
\begin{align}\label{eq:SineMode}
{\bf w}({\bf r})={\bf e}_z \sin\left(\frac{\pi z}{L_z}\right),
\end{align}
for $-L_z/2 \leq z \leq L_z/2$. The dimension of the cuboid is set by the effective mass and frequency of the oscillator, yielding for its ground mode $L_x\times L_y\times L_z \approx 0.31\,\mu{\rm m}\times0.31\,\mu{\rm m}\times0.84\,\mu{\rm m}$, using the speed of sound $v = 8433$\,m/s and density $\varrho = 2300$\,kg/m$^3$ of silicon.

This can now be used to calculate the Lindblad operators \eqref{eq:HarmOscFullLindbladOp}. 

The likelihood \eqref{eq:likenano} can be calculated with the characteristic function \eqref{eq:soldifflim} of the state \eqref{eq:rhopm} as a phase space integral
\begin{align}
P(\pm_1,\pm_2|\tau_e,\sigma,I)=\int d^2{\bf Q}d^2{\bf P}
\chi^{\pm_1}_t({\bf Q},{\bf P})\eta^{\pm_2}({\bf Q},{\bf P}),
\label{eq:MHOProbabilities}
\end{align} 
where $\eta^{\pm_2}({\bf Q},{\bf P})$ is the characteristic symbol of the operator \eqref{eq:MHORiedingerProjectorPositive}.

\begin{figure}
  \centering
  \includegraphics[width=0.45\textwidth]{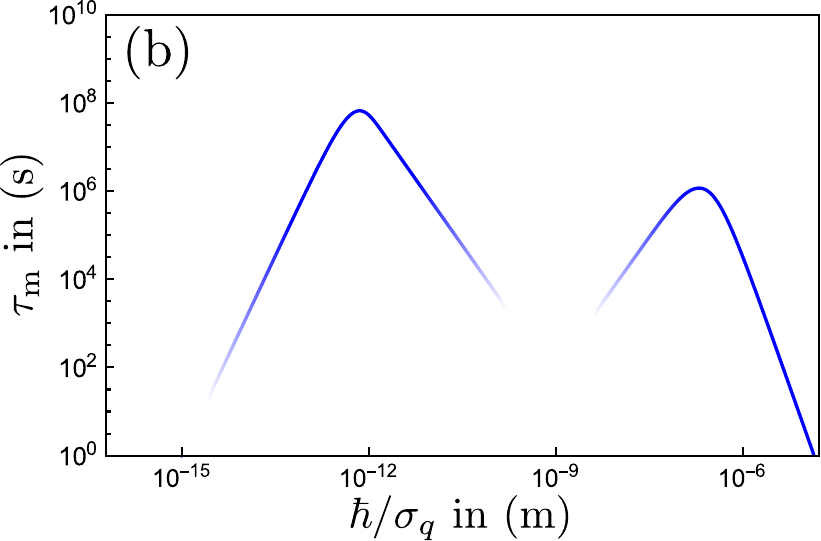}
  \caption{ The maximally excluded time parameter $\tau_{\rm m}$ as defined by the five percent quantile obtained via Bayesian updating with Eq.~\eqref{eq:Ppm222}. The local maximum to the right is assumed for  values of $\hbar/\sigma_q$ roughly equal to the spatial extension of the crystal mode $L_{x,z}$. The global maximum is achieved at $\hbar/\sigma_q\simeq\sqrt{\hbar^2/2m_{\rm Si}E_{\rm b}}$ where the momentum transfers become sufficiently strong to remove particles from the crystal. The fading of the graph indicates where the analytical descriptions derived in App.~\ref{app:b} fail: First, when $\hbar/\sigma_q$ is on the order of several \AA ngstr\"om so that the mass density can no longer be approximated as continuous, and second, when $\hbar/\sigma_q$ is on the order of femtometers where the diffusive regime ceases to be valid.}
\label{fig:BeyondContinuum}
\end{figure}

This expression can now be simplified by noting that the oscillator frequency is large on the timescale of the experiment, $\omega t \gg 1$, so that the time-averaged phase space coordinates \eqref{eq:MarmoOscFreeQP} can be used in the exponent of \eqref{eq:soldifflim},
\begin{align}
\chi_t({\bf Q},{\bf P}) \approx &
\exp\left[
-\frac{U_<(\sigma) t}{2 \tau_e} \left ( {\bf Q}^2 + \varrho^2 V_{\rm m}^2 (\Omega^{-1} {\bf P})^2 \right )\right]
\nonumber\\
&\times\chi_0({\bf Q}_t,{\bf P}_t)\,.
\end{align}
Moreover, the modification cannot create coherences between the oscillator states. In Eq.~\eqref{eq:MHORiedingerProjectorPositive} one can therefore keep only the diagonal terms and the initial coherences between ground state and first excited states,
\begin{align}
{\sf F}_{\pm_2} = \frac{1}{2} \left (\mathbb{1} - \left|0,0\right\rangle\left\langle 0,0\right|\pm e^{i\theta} \left|1,0\right\rangle\left\langle 0,1\right| \pm  e^{-i\theta}\left|0,1\right\rangle\left\langle 1,0 \right|\right) .
\label{eq:MHORiedingerProjectorPositive2}
\end{align}
The corresponding characteristic symbol is given in App.~\ref{app:b}, together with the characteristic function of the state \eqref{eq:rhopm}.

The integral Eq.~\eqref{eq:MHOProbabilities} yields the likelihood in its final form,
\begin{align}\label{eq:Ppm222}
&P(\pm_1,\pm_2|\tau_e,\sigma,I)=\nonumber\\
&\frac{1}{4\mathcal{N}}
+\frac{(\pm_1)(\pm_2)  4\cos (\theta- \Delta \Omega t )-2\xi t/\tau_{ e}-\xi^2 t^2/\tau^2_{ e}}{\mathcal{N}(2+\xi t/\tau_{ e})^4},
\end{align}
where $\Delta \Omega = 2\pi \times 45$\,MHz is the frequency mismatch between the oscillators, and we defined the dimensionless parameter $\xi = 2 U_<(\sigma)  \hbar/\varrho V_{\rm m} \omega$ characterizing the sensitivity of the relevant nanobeam mode to the modification parameter $\sigma_q$. The geometric factor  $U_<$, as defined in  Eq.~\eqref{eq:HarmOscDiffFT}, is evaluated in App.~\ref{app:b}.

The phase- or time-sweep measurement protocols performed in \cite{Riedinger2018} are described by varying $\theta$ and $t$, respectively. The (unreported) initial phase is deduced to be $ \phi\approx1.8\, {\rm rad}-\Delta\Omega\times123\,{\rm ns}$ by optimization. In order to obtain the achieved macroscopicity, we perform Bayesian updating to determine the posterior \eqref{eq:posterior} and maximize over $\sigma_q$. The resulting $\tau_{\rm m}$ is plotted in Fig.~\ref{fig:BeyondContinuum} for $q_c = \sqrt{2 m_{\rm Si} E_\mathrm{b}}$ with $E_{\rm b}=4.6\,$eV \cite{Farid1991}. It exhibits a global maximum of $\tau_{\rm m}=6.6\times 10^{7}\,$s at $\hbar/\sigma_q\simeq\sqrt{\hbar^2/2m_{\rm Si}E_{\rm b}}$,  yielding a macroscopicity value of $\mu_{\rm m}=7.8$.

Given the relatively high mass of the nanomechanical oscillators  and the fairly long coherence time achieved, one might expect the entangled nanobeams to be characterized by a higher degree of macroscopicity. That this is not the case can be traced back to the fact that the superposition state is delocalized only on the scale of femtometers. For such small spatial delocalizations, the sole influence of the modification is to add momentum diffusion to the nanobeam dynamics, leading to weakest possible form of spatial decoherence.

\section{ Conclusion}

The empirical measure discussed in this article serves to quantify the macroscopicity  reached in quantum mechanical superposition experiments by the degree to which they rule out classicalizing modifications of quantum theory. We showed how the framework of Bayesian hypothesis testing allows one to assess diverse experiments based on their raw data, thus accounting appropriately for all measurement uncertainties. The fact that measurement errors are fundamentally unavoidable, ensures that the macroscopicity $\mu_{\rm m}$ will always converge to a finite value, even if quantum mechanics holds on all scales. For sufficiently large data sets, when statistical errors tend to be  negligible, the here presented measure will approach the one given in Ref. \cite{Nimmrichter2013} for interferometric superposition tests. Equation \eqref{eq:Macroscopicity} is thus the natural generalization of the latter.

A great benefit of the formalism is that it allows one to straightforwardly combine independent parts of an experiment, e.g. quantum random walks of different lengths (Sec.\,\ref{sec:4}) or different measurement protocols for entangled nanobeams (Sec.\,\ref{sec:5}). Moreover, the Bayesian updating process naturally allows for correlated observables to be taken into account, as for instance the total atom number and the population imbalance in BEC interferometers (see Sec.\,\ref{sec:3}). Finally, the use of 
Jeffreys' prior ensures that the macroscopicity measure is solely determined by the experimental data at hand, irrespective of prior beliefs. In particular, using this  least informative prior prevents the macroscopicity measure to favor  any one type of quantum test against others. We showed that Jeffreys' prior exists for all physically relevant situations, where the likelihood is a smooth function of the modification parameters.

These advantages come at the cost that the required likelihoods are in general considerably more difficult to determine than e.g.\ specific coherences of the statistical operator. It requires one to capture appropriately how the relevant quantum  degrees of freedom are affected by the master equation \eqref{eq:modvonneum} describing the impact of the modification on the many-particle system state. 
We explained in Secs.~\ref{sec:3}-\ref{sec:5} how this works in practice for three rather different quantum superposition tests. 

We reemphasize that a naive application of the macroscopicity measure may yield a finite value even for experiments demonstrating no quantum superposition, because already the absence of observed heating can constrain the classicalization parameters. To be on the safe side, one must identify those observations that yield information only about modification-induced heating and use this data to condition the likelihoods as described at the end of Sect.~\ref{sec:2A}. In most quantum tests this is not necessary because the conditioning is already implemented in the measurement protocol.

The measure of macroscopicity put forward in this article 
can be used for any superposition test, provided a mechanical degree of freedom is involved, be it the electronic excitation of an atom or the motion of a kilogram-scale mirror. As such it does not apply to quantum tests involving only spins or photons. It seems natural to generalize the macroscopicity measure to pure photon experiments by drawing on a  minimal class of classicalizing modifications of QED, but it is  still an open problem how to get hold of the latter.
Beyond the assessment of macroscopicity, 
the Bayesian hypothesis testing presented in Sec.\,\ref{sec:2}, can also be used for a proper 
statistical description of tests of specific modification models, e.g.\ the various extensions of the  Continuous Spontaneous Localization model \cite{Bassi2013}, but also of environmental decoherence mechanisms.

Finally, it goes without saying that the macroscopicity $\mu_{\rm m}$ attributed to a given superposition test serves to highlight a single aspect of the experiment, albeit an important one. It must not be taken as a proxy for the overall significance of an experimental finding.

\acknowledgements

We thank Andrea Alberti, Tarik Berrada, and Ralf Riedinger for helpful comments on their experiments, and the authors of Ref.~\cite{Riedinger2018} for providing us with the  unpublished raw data reported in Fig.~\ref{fig:4}. BS thanks Gilles Kratzer for helpful discussions on the topic of Bayesian statistics. This work was funded by Deutsche Forschungsgemeinschaft (DFG, German Research Foundation) -- 298796255.

\appendix
\onecolumngrid

\section{Integrability of the posterior distribution} \label{app:beweis}

To see that Jeffreys' prior \eqref{eq:JeffreysPrior} always yields a normalizable posterior distribution \eqref{eq:posterior}, we first consider the limit $\tau_e \to \infty$, where the modification becomes arbitrarily weak. In this case  the general solution of the master equation \eqref{eq:modvonneum} can be expanded to first order in $1/\tau_e$ by its Dyson series. Calculating the likelihood then yields
\begin{align} \label{eq:exp}
P(D|\tau_e,\sigma,I) \simeq P_\infty(D |I)+\frac{1}{\tau_e} q(D|\sigma,I) \quad \text{for} \quad \tau_e \to \infty,
\end{align}
where $q$ is independent of $\tau_e$. Inserting the expansion \eqref{eq:exp} into Jeffreys' prior (\ref{eq:JeffreysPrior}) yields
\begin{align}
p(\tau_e | \sigma, I ) \stackrel{\tau_e\to\infty}{\sim} \begin{cases}
\tau_e^{-3/2} & \exists\, d_0 :P_\infty(d_0|I)=0, \\
\tau_e^{-2} & \, \text{else},
\end{cases}
\end{align}
implying that the posterior \eqref{eq:posterior} decays at least as $\tau_e^{-3/2}$ for $\tau_e \to \infty$.

Second, for $\tau_e \to 0$, where modification-induced decoherence and heating get stronger and stronger, we use that the likelihood $P(D|\tau_e,\sigma,I)$ will continuously approach some limiting classical probability,
\begin{equation}
P(D|\tau_e,\sigma,I) \simeq {P}_0(D|\sigma, I) + \tau_e^{\alpha} \tilde{q}(D|\sigma,I) \quad \text{for} \quad \tau_e \to 0,
\end{equation}
where $\alpha>0$ may depend on $D$. Using this to evaluate Jeffreys' prior \eqref{eq:JeffreysPrior} yields that
\begin{equation}
p(\tau_e | \sigma, I ) \stackrel{\tau_e\to 0}{\sim} \begin{cases}
\tau_e^{-(1 - \alpha_{\rm min}/2)} & \exists\, d_0 :P_0(d_0|\sigma,I)=0, \\
\tau_e^{-(1 - \alpha_{\rm min})} & \, \text{else},
\end{cases} 
\label{eq:LimitTauToZero}
\end{equation}
where $\alpha_{\rm min}>0$ is the minimal $\alpha$. Physically speaking, this means that no quantum superposition test will support a classical model of infinitely strong heating. Equation~\eqref{eq:LimitTauToZero} implies that the posterior always diverges weaker than $1/\tau_e$ for $\tau_e \to 0$.

Finally, to rule out that the posterior diverges at a finite $\tau_e \in (0,\infty)$, we note that the likelihood  $P(D|\tau_e,\sigma,I)$ stays non-negative for all $\tau_e$. Thus, whenever it vanishes for some value of $\tau_e$, its first derivative must also be zero and its second derivative must be non-negative. Application of  L'Hospital's rule then shows that the posterior stays finite for all intermediate values of $\tau_e$. This completes the argument why the choice of Jeffrey's prior \eqref{eq:JeffreysPrior} always leads to a normalizable posterior \eqref{eq:posterior} and thus yields a well-defined value of macroscopicity \eqref{eq:Macroscopicity}.

\section{Simultaneous shearing and diffusion of number squeezed BECs} \label{app:a}

For simultaneous phase diffusion and shearing the time evolution of the tangent space Wigner function $w_t(j_y,j_z)$ is given by the equation
\begin{align}
\partial_t w_t(j_y,j_z)=-\left(\frac{\epsilon}{\hbar}+2\zeta j_z\right)\partial_{j_y}w_t(j_y,j_z)+
\frac{\Gamma_{\rm P}}{2}\partial^2_{j_y} w_t(j_y,j_z)\,,
\end{align} 
which is solved by \eqref{eq:wignert}.
If diffusion takes place perpendicular to the shearing, the time evolution is given by the equation
\begin{align}
\partial_t w_t(j_y,j_z)=-2\zeta j_z\partial_{j_y}w_t(j_y,j_z)+
\frac{\Gamma_{\rm S}}{4}\partial^2_{j_z} w_t(j_y,j_z)\,,
\label{eq:ShearingDiffusion}
\end{align} 
without the free rotation around the general Bloch sphere that can be executed subsequently. Its general solution is 
\begin{align}
w_t(j_y,j_z)=\frac{1}{4\pi^2}\int dp_ydp_zdj_y'dj_z'w_0(j_y',j_z')e^{i(j_y-2\zeta j_z t)p_y-izp_z-ip_yj_y'+ip_z j_z'}\exp\left[-J^2\Gamma_{\rm S}\left(\frac{ p_z^2}{4}t-\frac{ \zeta p_yp_z}{2}t^2-\frac{\zeta^2 p_y^2}{3}t^3\right)\right]\,.
\end{align}
We take the initial distribution $w_0(j_y,j_z)$ to be a Gaussian with widths $\sigma_y$ and $\sigma_z$. Integrating  $j_z$ preserves the Gaussian form, yielding the marginal distribution
\begin{align}
w_t(j_y) = \int dj_z\,w_t(j_y,j_z)=\frac{1}{\sqrt{2\pi\sigma^2_y(t)}}\exp\left[\frac{j_y^2}{2\sigma^2_y(t)}\right],
\end{align} 
with variance
\begin{align}
\sigma^2_y(t)=\sigma^2_y+4\zeta^2t^2\left(\sigma_z^2+\frac{J^2\Gamma_{\rm L}t}{6}\right).
\end{align}

\section{Calculational details for the entangled nanobeams experiment} \label{app:b}
\subsection{Normalization of displacement fields}  

The equation of motion of a classical displacement field in an isotropic elastic medium can be derived from the Lagrangian density \cite{fetter2003theoretical}
\begin{align}
\mathcal{L}=\frac{\varrho}{2}\dot{\mathbf u}^2({\mathbf r},t) -V\left[{\bf u}({\bf r},t)\right] =\frac{\varrho}{2}\dot{\bf u}^2-
\frac{1}{4}\sum_{klmn=1}^3\left(\frac{\lambda}{2}\delta_{kl}\delta_{mn}+\mu \delta_{km}\delta_{ln}\right)\left(\frac{\partial u_k}{\partial x_l}+\frac{\partial u_l}{\partial x_k}\right)\left(\frac{\partial u_m}{\partial x_n}+\frac{\partial u_n}{\partial x_m}\right),
\end{align} 
where $\lambda$ and $\mu$ are the Lam\'e coefficients. Thus, the dynamics of ${\bf u}({\bf r},t)$ are given by 
\begin{align}
\varrho\frac{\partial^2{\bf u}}{\partial t^2} = \mu\nabla^2{\bf u}+(\lambda+\mu)\nabla(\nabla\cdot{\bf u}).
\end{align}
This equation can be solved by introducing the mode functions ${\bf u}_k({\bf r},t)$ as the eigenfunctions of the differential operator on the left hand side with eigenvalues $-\omega_k^2 \varrho$. The total displacement field can then be written as
\begin{align}
{\bf u}({\bf r},t)=\sum_k\sqrt{\frac{\hbar}{2 \rho V_k \omega_k}}{\bf w}_k({\bf r})\left(e^{-i\omega_k t}a_k+e^{i\omega_k t}a_k^*\right),
\label{eq:ClassicalDisplacementMode}
\end{align} 
so that its mean energy is
\begin{align}
\left\langle E\right\rangle_t=\left\langle \int d^3{\bf r} \left[
\frac{\varrho}{2}\dot{{\bf u}}^2({\bf r},t)+V\left[{\bf u}({\bf r},t)\right]
\right]\right\rangle_t
=\sum_k\frac{\hbar\omega_k}{V_k}\int d^3{\bf r}\, {\bf w}_k^2({\mathbf r})a_k^*a_k.
\label{eq:MHOmeanenergy}
\end{align}  
Demanding that $\left\langle E\right\rangle_t=\sum_k\hbar\omega_k a_k^*a_k$ yields the normalization condition $\int d^3{\mathbf r}\, {\mathbf w}_k^2({\mathbf r})=V_k$.

\subsection{Characteristic functions of mechanical oscillator states} 

The characteristic function of the initial oscillator state in \eqref{eq:rhopm} for $\phi=0$ can be calculated as
\begin{align}
\chi^{\pm_1}({\bf Q},{\bf P})
 =&\frac{1}{2}\exp\left[-\frac{1}{4\hbar\varrho V_{\rm m}}\left({\bf P}\cdot\Omega^{-1}{\bf P}+\varrho^2 V_{\rm m}^2{\bf Q}\cdot\Omega{\bf Q}\right)\right]
\nonumber\\
&\times\left(
1-\frac{1}{4\hbar\varrho V_{\rm m}}\left[\sum_{\lambda = 1,2}(\pm_1)^{\lambda}\left(\Omega^{-1/2}{\bf P}\right)_{\lambda}\right]^2
-\frac{\varrho V_{\rm m}}{4\hbar}\left[\sum_{\lambda = 1,2}(\pm_1)^{\lambda}\left(\Omega^{1/2}{\bf Q}\right)_{\lambda}\right]^2
\right)\,,
\end{align}
where ${\rm \Omega}={\rm diag}(\Omega_1,\Omega_2)$.
In a similar fashion, one obtains the characteristic symbols of the effect \eqref{eq:MHORiedingerProjectorPositive2} as
\begin{align}
\eta^{\pm_2}({\bf Q},{\bf P})
=&\frac{1}{2}\delta({\bf Q})\delta({\bf P})
-\left[1\pm_2\cos\theta\left(\frac{P_1P_2}{\hbar\varrho V_{\rm m}\sqrt{\Omega_1\Omega_2}}+Q_1Q_2\frac{\varrho V_{\rm m}\sqrt{\Omega_1\Omega_2}}{\hbar}\right)\pm_2\sin\theta\left(P_1Q_2\sqrt{\frac{\Omega_2}{\hbar^2\Omega_1}}-P_2Q_1\sqrt{\frac{\Omega_1}{\hbar^2\Omega_2}}\right)
\right]\nonumber\\
&\times\frac{1}{8\pi^2\hbar^2}\exp\left[-\frac{1}{4\hbar\varrho V_{\rm m}}\left({\bf P}\cdot\Omega^{-1}{\bf P}+\varrho^2 V_{\rm m}^2{\bf Q}\cdot\Omega{\bf Q}\right)\right].
\end{align}

\subsection{The geometric factor $U_<(\sigma)$} 
Assuming a continuous mass density, valid if $\hbar/\sigma_q\gg 5$\AA , the geometric factor \eqref{eq:HarmOscDiffFTParticleLoss} can be evaluated for the longitudinal mode \eqref{eq:SineMode} as
\begin{align}\label{eq:U}
U_<(\sigma)\simeq&U(\sigma)=\frac{2\varrho^2\hbar^7}{m_e^2\sigma_q^7L_z^3}\left(1-e^{-L_x^2\sigma_q^2/2\hbar^2}+\frac{\sqrt{\pi}L_x\sigma_q}{\sqrt{2}\hbar}{\rm erf}\left[\frac{L_x\sigma_q}{\sqrt{2}\hbar}\right]\right)^2\nonumber\\
&\times\left[\sqrt{2\pi}\left(h\left[\frac{L_z\sigma_q}{\hbar},0\right]+e^{-L_z^2\sigma_q^2/2\hbar^2}{\rm Re}\left\{h\left[\frac{L_z\sigma_q}{\hbar},\frac{L_z\sigma_q}{\hbar}\right]\right\}\right)-\left(1+e^{-L_z^2\sigma_q^2/2\hbar^2}\right)\frac{L_z\sigma_q}{\hbar}\left(\pi^2-2\frac{L_z^2\sigma_q^2}{\hbar^2}\right)\right],
\end{align}
with
\begin{align}
h[a,b]=\sqrt{\frac{\pi}{2}}\left(i3a^2+\pi b^2-i\pi^2\right)\exp\left[\frac{(\pi/a-ib)^2}{2}\right]{\rm erf}\left[\frac{i \pi/a+b}{\sqrt{2}}\right].
\end{align}

If $\hbar/\sigma_q$ is on the order of the lattice constant, the approximation of a continuous mass density fails. For even smaller $\hbar/\sigma_q$ the Gaussian in \eqref{eq:HarmOscDiffFT} suppresses all contributions involving more than a single atom, so that the modification acts on each of the $N$ atoms individually. The geometric factor then reads as
\begin{align}
U_<(\sigma)=N\frac{m_{\rm Si}^2}{4\hbar^2m_e^2}{\rm erf}\left(\frac{q_c}{\sqrt{2}\sigma_q}\right)^2\left[\sigma_q^2{\rm erf}\left(\frac{q_c}{\sqrt{2}\sigma_q}\right)^2-\sqrt{\frac{2}{\pi}}\sigma_q q_c e^{-q_c^2/2\sigma_q^2}\right].
\end{align}
Here, we averaged the mode function \eqref{eq:SineMode} over the whole crystal, $\sum_n{\mathbf w}^2({\mathbf r}_n)\approx N/2$. As a result, the diffusion increases quadratically with $\sigma_q$ until the momentum displacements are strong enough to remove the particles from the crystal. In the limit that $\sigma_q\ll q_c$ one obtains $U_<(\sigma) \simeq Nm_{\rm Si}^2\sigma_q^2/4\hbar^2m_e^2$.

\twocolumngrid

\end{document}